\documentclass[a4paper]{revtex4}
\usepackage{epsfig,amsmath,amssymb}
\def\Qz{{\bf Q}}
\def\qj{{\bf q}_j}
\def\qi{{\bf q}_i}
\def\qk{{\bf q}_k}
\def\tpm{$\pm$~}
\def\q{q \bar q}
\def\gs{\gamma_s}
\def\epe{e$^+$e$^-$~}
\def\eeqq{$e^+e^- \to q \bar q$}

\def\eecc{$e^+e^- \to c \bar c$}
\def\eebb{$e^+e^- \to b \bar b$}

\def\e{{\rm e}}
\def\i{{\rm i}}
\def\d{{\rm d}}
\def\B{\boldmath}
\def\sqrtsnn{$\sqrt{s}$}

\def\be{\begin{equation}}
\def\ee{\end{equation}}

\def\NP{{ Nucl.\ Phys.\ }}
\def\PL{{ Phys.\ Lett.\ }}
\def\PR{{ Phys.\ Rev.\ }}

\def\PRL{{ Phys.\ Rev.\ Lett.\ }}

\def\ZP{{ Z.\ Phys.\ }}
\def\EP{{ Eur.\ Phys.\ J.\ C}}

\begin{document}

\title{The Thermal Production of Strange and Non-Strange Hadrons
in \B\epe Collisions} 

\author{F. Becattini}
\affiliation{Dipartimento di Fisica, Universit\`a di Firenze, and INFN Sezione di Firenze} 
\author{P. Castorina}
\affiliation{Dipartimento di Fisica, Universit{\`a} di Catania, and INFN Sezione di Catania}
\author{J. Manninen}
\affiliation{INFN Sezione di Firenze} 
\author{H. Satz}
\affiliation{Fakult\"at f\"ur Physik, Universit\"at Bielefeld}

\begin{abstract}
The thermal multihadron production observed in different high energy 
collisions poses two basic problems: (1) why do even elementary 
collisions with comparatively few secondaries (\epe annihilation) 
show thermal behaviour, and (2) why is there in such interactions a 
suppression of strange particle production? We show
that the recently proposed mechanism of thermal hadron production through 
Hawking-Unruh radiation can naturally account for both. The event horizon 
of colour confinement leads to thermal behaviour, but the resulting 
temperature depends on the strange quark content of the produced hadrons, 
causing a deviation from full equilibrium and hence a suppression of 
strange particle production. We apply the resulting formalism to 
multihadron production in $e^+e^-$ annihilation over a wide energy range 
and make a comprehensive analysis of the data in the conventional 
statistical hadronization model and the modified Hawking-Unruh formulation. 
We show that this formulation
provides a very good description of the measured hadronic abundances, fully 
determined in terms of the string tension and the bare strange quark mass; 
it contains no adjustable parameters. 
\end{abstract}

\maketitle

\section{Introduction}

\begin{figure}[h]
\vspace*{-0.5cm}
\centerline{\psfig{file=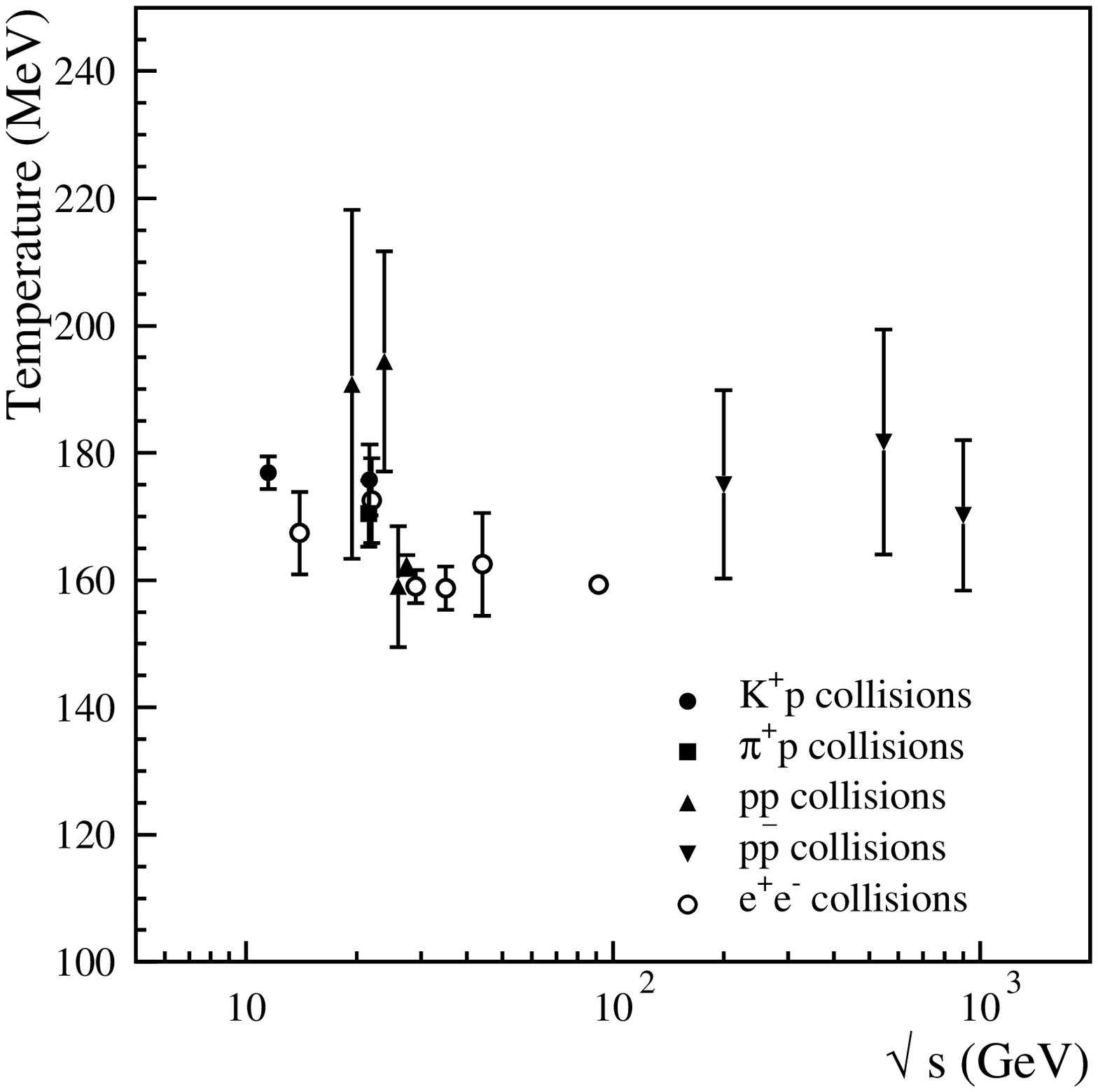,height=9.5cm}} 
\caption{Hadronization temperatures (from ref.~\cite{Beca-Biele}).}
\label{temp}
\end{figure}

Hadron production in high energy collisions shows remarkably universal thermal 
features. In $e^+e^-$ annihilation \cite{Beca-e,erice,Beca-h}, in $pp$, 
$p\bar p$ \cite{Beca-p} and more general $hh$ interactions \cite{Beca-h}, 
as well as in the collisions of heavy nuclei \cite{Beca-hi}, over an energy range from
around 10 GeV up to the TeV range, the relative abundances of the produced 
hadrons appear to be those of an ideal hadronic resonance gas at a quite
universal temperature $T_H \approx 160-170$ MeV, as illustrated in Fig.\ 
\ref{temp} \cite{Beca-Biele}. The transverse momentum spectra of the hadrons 
produced in hadronic collisions at intermediate energies are also in good
agreement with the predictions of a statistical model based on the same
temperature \cite{Beca-h}.

\medskip

There is, however, one important non-equilibrium effect observed:
the production of strange hadrons in elementary collisions is suppressed 
relative to an overall equilibrium. This is usually taken into account 
phenomenologically by introducing an overall strangeness suppression factor 
 $\gs < 1$ \cite{Raf}, which reduces the predicted abundances by $\gs, 
\gs^2$ and $\gs^3$ for hadrons containing one, two or three strange quarks 
(or antiquarks), respectively. In high energy heavy ion collisions, 
strangeness suppression becomes less and disappears at high energies 
\cite{gamma}.

\medskip

When comparing the temperatures of different collision channels in Fig.\ 
\ref{temp}, it should be stressed that in elementary collisions, in contrast 
to heavy ion collisions, the conservation of charge, baryon number and 
strangeness is enforced exactly (canonical ensemble), so that here no chemical 
potential is introduced. The effect of exact charge conservation is important 
in elementary collisions because of the low multiplicities involved, while it 
is generally negligible in large multiplicity high energy nuclear reactions.
The fitted values of $\gs$ lie around 0.5 to 0.7 in elementary collisions; 
the heavy ion values appear to vary as a function of energy, but it tend to 
approach unity \cite{gamma}.

\medskip

The origin of the observed thermal behaviour has been an enigma for many years 
and there is a still ongoing debate about the interpretation of these
results \cite{diba}. While the common belief is that in high energy heavy ion 
collisions multiple parton scattering could lead to kinetic thermalization 
through multiple scattering, $e^+e^-$ or elementary hadron interactions do 
not readily 
allow such a description. The universality of the observed temperatures, on
the other hand, suggests a common origin for all high energy collisions, 
and it was recently proposed \cite{CKS} that thermal hadron production is the 
QCD counterpart of Hawking-Unruh radiation \cite{Hawk,Un}, emitted at the
event horizon due to colour confinement. A well-known instance of this
phenomenon is the Schwinger mechanism \cite{Schwinger,KT} of pair production 
in a constant electric field $\cal E$. The probability of 
spontaneously producing an electron-positron pair is in this case given by
\be
P(m.{\cal E}) \sim \exp\{-\pi m^2/e{\cal E}\},
\ee
where $m$ is the electron mass and $e$ its charge. Since 
\be
a_e= {2~\!e{ \cal E} \over m}
\ee
is just the acceleration of an electron (of reduced mass $m/2$)
in the field $\cal E$, we find that
the pair production probability has the thermal form
\be
P(m,{\cal E}) \sim \exp\{-m/T_U\},
\ee
where
\be
T_U = {a_e \over 2 \pi} = {e{\cal E} \over \pi m}
\ee
is the corresponding temperature. In other words, it is just that 
obtained by Unruh \cite{Un} for the radiation emitted when a 
mass suffers constant acceleration and hence encounters an event horizon. 

\medskip

In the case of QCD and approximately massless quarks, the resulting 
hadronization temperature is determined by the string tension $\sigma$, with 
$T \simeq \sqrt{\sigma/2\pi}$. The aim of the present work is to show that 
strangeness suppression occurs naturally in this framework, without 
requiring a specific suppression factor. The crucial role here is played 
by the non-negligible strange quark mass, which modifies the emission 
temperature for such quarks. 

\medskip

In this work, we will focus on \epe collisions, which is the simplest case for 
this model to be tested. In the next section, we will briefly review the usual 
formulation of the statistical hadronization model, since this will form the
basis also for the subsequent description in terms of Hawking-Unruh radiation 
model, to be formulated in Sect.~3. There we shall in particular derive the 
dependence of the radiation temperature on the mass of the produced quark
and show how this affects the hadronization in $e^+e^-$ annihilation. 
In Sect.~4, we then present an updated comprehensive analysis of all available 
data from $\sqrt s = 14$ GeV to $\sqrt s = 189$ GeV and compare the 
results of the Hawking-Unruh formulation to the conventional statistical
description.

\section{Statistical Hadronization in ${\rm e}^+{\rm e}^-$ Collisions}

In this section we will briefly review the essentials of the statistical 
hadronization model and its application to \epe collisions. For a detailed
description see ref.~\cite{meaning}.

\medskip

The statistical hadronization model assumes that hadronization in high 
energy collisions is a universal process proceeding through the formation of 
multiple colourless massive clusters (or fireballs) of finite spacial 
extension. These clusters are taken to decay into hadrons according 
to a purely 
statistical law: every multi-hadron state of the cluster phase space 
defined by its mass, volume and charges is equally probable.
The mass distribution and the distribution of charges (electric, baryonic 
and strange) among the clusters and their (fluctuating) number are 
determined in the prior dynamical stage of the process. 
Once these distributions are known, each cluster can be hadronized on
the basis of statistical equilibrium, leading to the calculation of 
averages in the {\em microcanonical ensemble}, enforcing the exact 
conservation of energy and charges of each cluster. 

\medskip

Hence, in principle, one would need the mentioned dynamical information 
in order to make definite quantitative predictions to be compared with 
data. Nevertheless, for Lorentz-invariant quantities such as multiplicities, 
one can introduce a simplifying assumption and thereby obtain a simple 
analytical expression in terms of a temperature. The key point is to assume
that the distribution of masses and charges among clusters is again
purely statistical \cite{Beca-h}, so that, as far as the calculation 
of multiplicities is concerned, the set of clusters becomes equivalent, 
on average, to a large cluster ({\em equivalent global cluster}) whose 
volume is the sum of proper cluster volumes and whose charge is the sum of 
cluster charges (and thus the conserved charge of the initial colliding 
system). In such a global averaging process, the equivalent cluster 
generally turns out to be large enough in mass and volume so that the 
canonical ensemble becomes a good approximation of the more fundamental 
microcanonical ensemble \cite{Beca-micro}; in other words, a temperature 
can be introduced which replaces the {\sl a priori} more fundamental 
description in terms of an energy density. 

\medskip

It should be stressed that in such an analysis of multiplicities, temperature 
has essentially a global meaning and not local as in hydrodynamical models. 
The only local meaningful quantity is cluster's energy density, and even 
though the globally fitted temperature value is closely related to it, this 
does not mean that single physical clusters can be described in terms of a 
temperature, unless they are sufficiently large (about 10 GeV in mass, see
\cite{Beca-micro}).

\medskip

In the statistical hadronization model supplemented by global cluster 
averaging, the primary multiplicity of each hadron species $j$ is given 
by \cite{Beca-h}
\be\label{mult}
 \langle n_j \rangle^{\rm primary} = \frac{V T (2J_j+1)}{2\pi^2} 
 \sum_{n=1}^\infty \gs^{N_s n}(\mp 1)^{n+1}\;\frac{m_j^2}{n}\;
 {\rm K}_2\left(\frac{n m_j}{T}\right)\, \frac{Z(\Qz-n\qj)}{Z(\Qz)}
\ee
where $V$ is the (mean) volume and $T$ the temperature of the equivalent 
global cluster. Here $Z(\Qz)$ is the canonical partition function depending 
on the initial abelian charges $\Qz = (Q,N,S,C,B)$, i.e., electric charge, 
baryon number, strangeness, charm and beauty, respectively. We denote by
$m_j$ and $J_j$ the mass and the spin of the hadron $j$, and 
$\qj = (Q_j,N_j,S_j,C_j,B_j)$ its corresponding charges; $\gs$ is the extra 
phenomenological factor implementing a suppression of hadrons containing 
$N_s$ strange valence quarks (see Sect.~1). In the sum (\ref{mult}), the upper 
sign applies to bosons and the lower sign to fermions. For temperature 
values of 160 MeV or higher, Boltzmann statistics corresponding to the 
term $n=1$ only in the series (\ref{mult}) is a very good approximation for 
all hadrons (within 1.5\%) but pions. For resonances, the formula (\ref{mult}) 
is folded with a relativistic Breit-Wigner distribution of the mass $m_j$. 

\medskip

The canonical partition function can be expressed as a multi-dimensional 
integral
\begin{eqnarray}\label{mpf}
 && Z(\Qz) = \frac{1}{(2\pi)^N} \int_{-\pi}^{+\pi} \, \d^N \phi \,\, 
 \e^{\i, \Qz \cdot \phi} \nonumber  \\ 
 && \times \exp \, \left[ \frac{V}{(2\pi)^3} \sum_j (2J_j+1) 
 \int \d^3 p \,\, \log \, (1 \pm \gamma_s^{N_{sj}} 
  \e^{-\sqrt{p^2+m_j^2}/T_i -\i \qj \cdot \phi})^{\pm 1} \right] \; .
\end{eqnarray}   
where $N$ is the number of conserved abelian charges. Unlike the 
grand-canonical case, the logarithm of the canonical partition function 
does not scale linearly with the volume. 
Therefore, the chemical factors $Z(\Qz-n\qj)/Z(\Qz)$ turn out to be less
than unity at finite volume (canonical suppression) and only asymptotically 
reach their grand-canonical limit of unity, for an initially completely 
neutral system. Indeed, they play a major role in determining particle 
yields in \epe collisions.

\medskip

For all energies considered here, the production of heavy flavoured 
hadrons is negligible in \eeqq~events, where $q$ is a light quark 
($u,d,s$). As a result, in formula (\ref{mult}) the charm and bottom charge 
can be neglected, so that the abelian charge vector reduces to a three-component 
form $\Qz = (Q,N,S)$ and can be calculated with a numerical integration 
\cite{Beca-p,chemfact}. On the other hand, 
\eeqq~ events, where $q$ is a heavy quark, always result in the production 
of two open heavy flavoured hadrons, arising from the primary heavy 
quark-antiquark pair, which subsequently decay into light-flavoured hadrons. 
In the statistical model, this constraint is readily 
implemented, requiring the number of heavy quarks plus antiquarks in the 
global description to be two; because of the high charm/bottom mass compared 
to the typical temperature value of 160 MeV, the probability of producing 
extra heavy $\q$ pairs is absolutely negligible. The multiplicities of 
charm (bottom) hadrons in \eecc~(\eebb) events become in the Boltzmann 
approximation \cite{Beca-p,erice}
\begin{equation}\label{hfmult}
 \langle n_j \rangle\
  = \gs^{N_{Sj}} z_j \,\, \frac{\sum_i \gs^{N_{si}}z_i \zeta(\Qz-\qj-\qi)}
   {\sum_{i,k} \gs^{N_{si}} \gs^{N_{sk}} z_i z_k \zeta(\Qz-\qi-\qk)} \; ,
\end{equation}
where
\be
 z_j = \frac{V}{2\pi^2} (2J_j + 1) {m_j^2 T} {\rm K}_2\left(\frac{m_j}{T}
 \right),
\ee
and $\zeta$ denotes the canonical partition function as in (\ref{mpf}),
involving only light-flavoured particles, with $N=3$. The indices 
$j,k$ label charm (bottom) hadrons, while the index $i$ their antiparticles. 

\medskip

The statistical treatment of heavy quark formation and hadronization
as outlined here effectively means heavy quarks occur only in 
\eeqq~interactions, leading to one open charm (bottom) hadron and
one corresponding antihadron per event. The relative rates of the
different possible open charm (bottom) states, however, are determined
by their phase space weights.

\section{String Breaking and Event Horizons}

\begin{figure}[h]
\centerline{\psfig{file=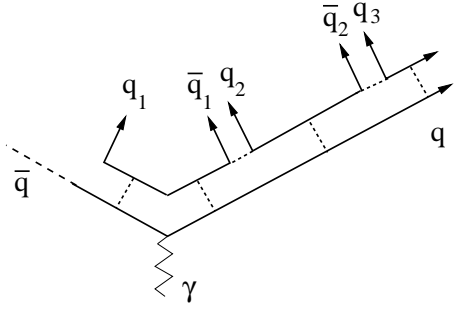,width=6cm}} 
\caption{String breaking through $\q$ pair production}
\label{breaking}
\end{figure}

We first outline the thermal hadron production process through
Hawking-Unruh radiation for the specific case of $e^+e^-$ annihilation
(see Fig.\ \ref{breaking}). The separating primary $\q$ pair excites a 
further pair $q_1\bar q_1$ from the vacuum, and this pair is in turn pulled 
apart by the primary constituents. In the process, the $\bar q_1$ shields 
the $q$ from its original partner $\bar q$, with a new $q\bar q_1$ string 
formed. When it is stretched to reach the pair production threshold, a 
further pair is formed, and so on \cite{
bj,nus}. Such a pair 
production mechanism is a special case of Hawking-Unruh radiation \cite{KT}, 
emitted as hadron $\bar q_1 q_2$  when the quark $q_1$ tunnels through its 
event horizon to become $\bar q_2$.  
The corresponding hadron radiation temperature is given by the Unruh form 
$T_H = {a/2 \pi}$, where $a$ is the acceleration suffered by the quark 
$\bar q_1$ due to the force of the string attaching it to the primary quark 
$Q$. This is equivalent to that suffered by quark $q_2$ due to the 
effective force of the primary antiquark $\bar Q$. Hence we have
\be
a_q = {\sigma \over w_q} =
{\sigma \over \sqrt{m_q^2 + k_q^2}}, 
\ee
where $w_q =\sqrt{m_q^2 + k_q^2}$ is the effective mass of the produced 
quark, with 
$m_q$ for the bare quark mass and $k_q$ the quark momentum inside the 
hadronic system 
$q_1\bar q_1$ or $q_2\bar q_2$. Since the string breaks \cite{CKS} when 
it reaches 
a separation distance 
\be
x_q \simeq {2\over \sigma} \sqrt{m^2_q + (\pi \sigma /2)},
\ee
the uncertainty relation gives us with $k_q \simeq 1/x_q$
\be
w_q  = \sqrt{m_q^2 + [\sigma^2/(4m_q^2 + 2\pi \sigma)]} 
\ee
for the effective mass of the quark. The resulting quark-mass dependent 
Unruh temperature is thus given by
\be
T(qq) \simeq {\sigma \over 2\pi w_q}.
\label{Tq}
\ee
Note that here it is assumed that the quark masses for $q_1$ and $q_2$
are equal. For $m_q \simeq 0$, eq.\ (\ref{Tq}) reduces to
\be\label{T0}
T(00) \simeq \sqrt{\sigma \over 2\pi},
\ee
as obtained in \cite{CKS}.

\medskip

If the produced hadron ${\bar q}_1 q_2$ consists of quarks of different 
masses, the resulting temperature has to be calculated as an average
of the different accelerations involved. For one massless quark 
($m_q \simeq 0$) and one of strange quark mass $m_s$, the average
acceleration becomes
\be
\bar a_{0 s} = {w_0 a_0 + w_s a_s \over w_0 + w_s} = 
{2\sigma \over w_0 + w_s}.
\ee
From this the Unruh temperature of a strange meson is given by
\be\label{T0s}
T(0s) \simeq {\sigma \over \pi (w_0 + w_s)},
\ee
with $w_0 \simeq \sqrt{1/2\pi\sigma}$. Similarly, we obtain
\be\label{Tss}
T(ss) \simeq {\sigma \over 2 \pi w_s},
\ee
for the temperature of a meson consisting of a strange quark-antiquark pair 
($\phi$).  
With $\sigma\simeq 0.2$ GeV$^2$, eq.\ (\ref{T0}) gives $T_0 \simeq 0.178$ 
GeV. A strange quark mass of 0.1 GeV reduces this to $T(0s) \simeq 0.167$ 
GeV and $T(ss)\simeq 157$ MeV, i.e., by about 6 \% and 12 \%, respectively.

\medskip

\begin{figure}[h]
\vskip-5.7cm
\hskip8cm{\psfig{file=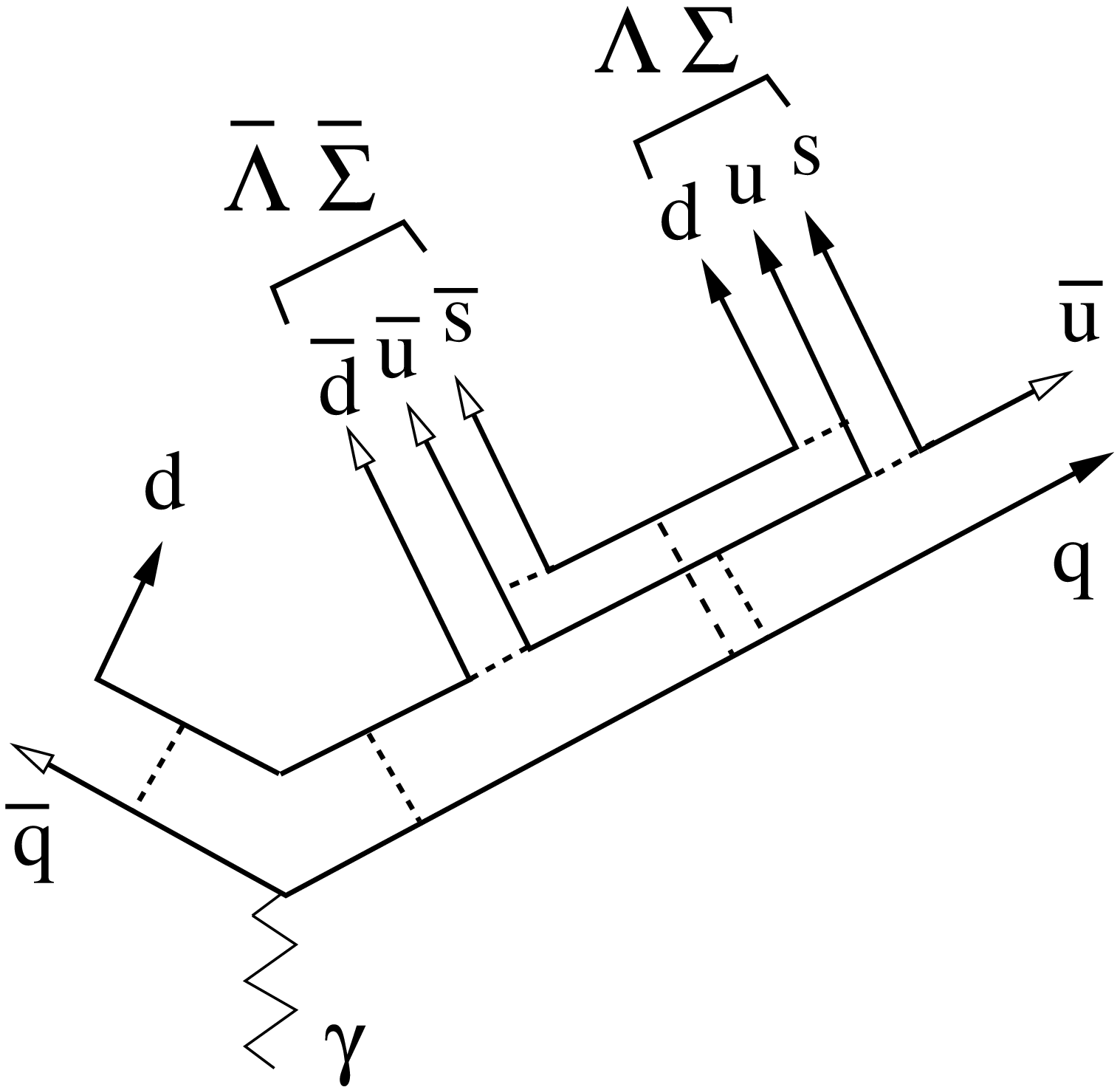,width=5.5cm}}
\bigskip
\caption{Nucleon (left) and hyperon (right) production in $e^+e^-$ 
annihilation}
\label{baryon}
\end{figure}

The scheme is readily generalized to baryons. The production
pattern is illustrated in Fig.\ \ref{baryon} and leads to an
average of the accelerations of the quarks involved. We thus have
\be
T(000) = T(0) \simeq {\sigma \over 2\pi w_0}
\ee
for nucleons, 
\be
T(00s) \simeq {3 \sigma \over 2\pi(2w_0 + w_s)}
\ee
for $\Lambda$ and $\Sigma$ production, 
\be
T(0ss) \simeq {3 \sigma \over 2\pi(w_0 + 2w_s)}
\ee
for $\Xi$ production, and 
\be
T(sss) = T(ss) \simeq {\sigma \over 2\pi w_s}
\ee
for that of $\Omega$'s. 
We thus obtain a resonance gas picture with five different hadronization 
temperatures, as specified by the strangeness content of the hadron in
question: $T(00)=T(000),~T(0s),~T(ss)=T(sss),~T(00s)$ and $T(0ss)$.

\begin{figure}[h]
\hskip-7cm{\psfig{file=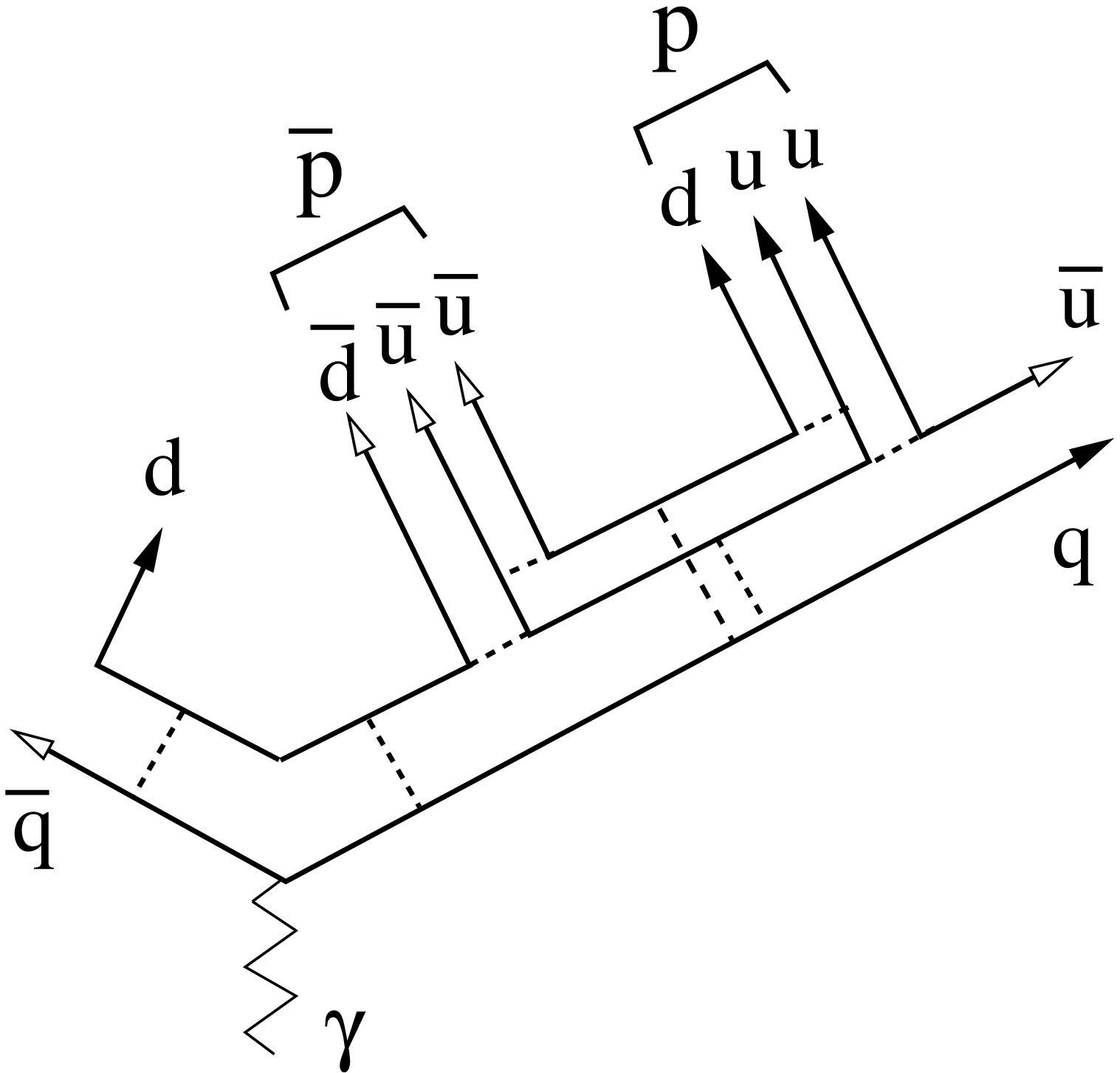,width=5.5cm}}
\end{figure}

\medskip

It is important to stress that, in this picture, the primary quarks produced 
directly in the annihilation are not related to Hawking-Unruh radiation, nor 
are the light quarks with which they eventually combine to hadronize. 
Therefore, leading hadrons (those containing primary quarks) are essentially 
different and should be treated separately from hadrons containing only 
newly produced quarks. In practice, while in the conventional statistical 
model the same hadronization temperature governs the relative probabilities 
of emitting different species of leading hadrons, in the Hawking-Unruh 
formulation, we do not have any specific prescription for this. The problem of 
calculating leading hadron yields is relevant in \epe~annihilations because, 
in contrast to hadronic collisions, the primary heavy quarks $c$ and $b$ 
have large branching ratios and significantly contribute to the production 
of light-flavoured hadrons through the decay chain, especially in the 
strange sector.
Lacking a definite recipe, we chose to calculate the relative heavy flavoured 
hadron yields by using the same temperatures as for light-flavoured ones,
quoted above, keeping one weight $w_0$ fixed and using $w_0$ or $w_s$ 
according to whether the heavy quark hadronization occurs through combination 
with either $u,d$ or with $s$, respectively. It should be noted out that this is 
not the only option and that different choices may lead to different results.

\medskip

The different species-dependent temperatures are to be inserted into the 
formulae (\ref{mult}) and (\ref{hfmult}) of the previous section, in order to 
determine the primary hadron multiplicities.
We note at this point a subtle conceptual difference between the conventional
statistical approach and the Hawking-Unruh formulation. The usual 
statistical description employs, as noted above, a global cluster
averaging, with each cluster statistically composed. In the Hawking-Unruh
scheme, the radiation in each step is a hadron formed from the emitted
$\q$ pair, not some thermal cluster. Since the hadron can, however,
be a highly excited resonance, the two descriptions become equivalent
in a Hagedorn-type picture proposing resonances made up of resonances in
a self-similar pattern.  

\medskip

The multiplicities obtained in the Hawking-Unruh scheme are, as emphasized, 
fully determined by the two basic parameters of the formulation, 
the string tension $\sigma$ and the strange quark mass $m_s$. Apart from 
possible variations of these quantities, our description is thus 
parameter-free. As illustration, we show in table \ref{tab:1} the 
temperatures obtained for $\sigma = 0.2$ GeV$^2$ and three different 
strange quark masses. It is seen that in all cases, the temperature for 
a hadron carrying non-zero strangeness is lower than that of non-strange 
hadrons; as we shall show, this leads to an overall strangeness suppression. 

\begin{table}[h]
\begin{center} 
\begin{tabular}{|c|c|c|c|}  \hline
\hline
$ T  $ & $m_s =0.075$ & $ m_s=0.100 $ & $m_s= 0.125 $ \\
\hline
$T(00)$ & 0.178  & 0.178  &  0.178 \\
$T(0s)$ & 0.172  & 0.167  &  0.162 \\
$T(ss)$ & 0.166  & 0.157  &  0.148 \\
$T(000)$& 0.178  & 0.178  &  0.178 \\
$T(00s)$& 0.174  & 0.171  &  0.167 \\
$T(0ss)$& 0.170  & 0.164  &  0.157 \\
$T(sss)$& 0.166  & 0.157  &  0.148 \\
\hline
\end{tabular}
\caption{Hadronization temperatures for hadrons of different strangeness 
content, for $m_s= 0.075 , 0.100,0.125$ Gev and $\sigma=0.2$ GeV$^2$.}
\label{tab:1}
\end{center}
\end{table}

\medskip

Our picture implies that the produced hadrons are emitted slightly ``out of 
equilibrium'', in the sense that the emission temperatures are not identical. 
As long as there is no final state interference between the produced quarks 
or hadrons, we expect to observe this difference and hence a modification 
of the production of strange hadrons, in comparison to the corresponding 
full equilibrium values. 
Once such interference becomes likely, as in high energy heavy ion
collisions, equilibrium can be at least partially restored, weakening
the strangeness suppression. The extension of our approach to heavy
ion collisions will be dealt with in a subsequent paper.

\section{Analysis of Hadron Multiplicities}

Multihadron production in \epe~annihilation has been studied at PEP, PETRA 
and LEP over an energy range from 14 to 189 GeV, and the multiplicities of a 
large number of different species have been measured. The relevant data
and their references are compiled in the Appendix.

\medskip

In order to compare the models with the experimental data, we have at each
energy made a fit to available measured multiplicities of light-flavoured 
hadrons, both in the traditional statistical model and in the Hawking-Unruh 
formulation. The traditional model has three free parameters to be 
determined, namely the temperature $T$, the global volume $V$ (see the 
discussion in the Sect.~2), and the strangeness undersaturation parameter 
$\gs$. In the Hawking-Unruh model, $V$ is kept, but the string tension 
$\sigma$ and the strange mass $m_s$ replace $T$ and $\gs$ as fit 
parameters.

\medskip

In the fit, the theoretical multiplicity of a given species, to be compared 
to the data, is calculated as the sum of primary multiplicity given by 
(\ref{mult}) and the contribution from the decay of unstable heavier hadrons,
\begin{equation}\label{branching}
\langle n_j \rangle  = \langle n_i \rangle^{\mathrm{primary}} + 
\sum_k \mathrm{Br}(k\rightarrow j) \langle n_k \rangle,
\end{equation}
where the branching ratios are the measured values as listed in the latest 
compilation of the Review of Particle Physics \cite{pdg}. For the decays 
of heavy flavoured hadrons with unknown branching fractions, we have used 
the predictions of the PYTHIA \cite{pythia} program. The hadrons 
considered unstable in \epe experiments are all species except 
$\pi$, $K^{\pm}$, $K^0_L$, $p$, $n$, and we have followed this convention 
in the theoretical calculation to meet the definition of measured 
multiplicities. The hadrons and resonances contributing 
to the sum in Eq.~(\ref{branching}) consist here of all known states 
\cite{pdg} up to a mass of 1.8 GeV.

\medskip

A specific fraction of \epe annihilations occurs into heavy $c$ and $b$ 
quarks. In this case, the multiplicities of light-flavoured hadrons are 
affected by the presence of the heavy quarks, both at primary level as 
the canonical partition function changes (see discussion in the previous 
section) and at final level because of the heavy flavoured hadron decays. 
This is taken into account in our calculations, and the production rate of 
the $j^{th}$ hadron is given by
\begin{equation}
\langle\!\langle n_j \rangle\!\rangle = 
\sum_{i=1}^5 R_i ~\langle n_j \rangle _{q_i}    
\end{equation}
where $i=1,...,5$ accounts for $u,d,s,c,b$ quarks and the index $q_i$ 
specifies the corresponding multiplicity in $e^+ e^- \to q_i \bar q_i$ 
annihilation. The values of  
\begin{equation}
 R_i = \frac{\sigma({\rm e}^+{\rm e}^-\rightarrow{\rm q}_i\overline{\rm q}_i)}
    {\sigma({\rm e}^+{\rm e}^-\rightarrow{\rm hadrons})}
\end{equation} 
are the corresponding branching fractions, obtained at each centre-of-mass
energy from measurements and electroweak calculations. We have here taken 
the values calculated in the PYTHIA programme \cite{pythia}, which are quoted 
in table \ref{bratios} for each centre-of-mass energy. 

\begin{table}[h]
\begin{center} 
\begin{tabular}{|c|c|c|c|c|}  \hline
\hline
  $\sqrt{s}$ & $R_u+R_d$ & $R_s$ & $R_c$ & $R_b$  \\
\hline
14           &  0.46	 &  0.09 & 0.37  &  0.08   \\
22           &  0.46	 &  0.09 & 0.36  &  0.09   \\
29           &  0.46	 &  0.09 & 0.36  &  0.09   \\
35           &  0.46	 &  0.09 & 0.36  &  0.09   \\
43           &  0.46  	 &  0.09 & 0.36  &  0.09   \\
91.25        &  0.39	 &  0.22 & 0.17  &  0.22   \\
133          &  0.41	 &  0.18 & 0.23  &  0.18   \\
161          &  0.42	 &  0.17 & 0.24  &  0.17   \\
183          &  0.42	 &  0.16 & 0.26  &  0.16   \\
189          &  0.42	 &  0.16 & 0.26  &  0.16   \\
\hline
\end{tabular}
\caption{Branching ratios for the \eeqq~annihilations into various
quark flavours as a function of centre-of-mass energy.}
\label{bratios}
\end{center}
\end{table}

The mass of resonances with $\Gamma > 1$ MeV has been distributed according 
to a relativistic Breit-Wigner function over the interval $[m_0-\Delta m, 
m_0+\Delta m]$, $m_0$ where $m_0$ is the central mass value and $\Delta m = 
\min \{2\Gamma, m_0-m_{th}\}$, $\Gamma$ being the width and $m_{th}$ the 
threshold mass value for the opening of all allowed decay channels. The 
primary production rate of neutral mesons such as $\eta$, $\eta'$, 
$\phi$, $\omega$ and others, which are a superposition of $s\bar s$ and 
$u \bar u$, $d \bar d$ states, has been suppressed with $\gs^2$ according 
to the fraction of $s\bar s$ content. For this purpose we use the mixing 
formulas quoted in the Review of Particles Properties \cite{pdg} with 
angles $\theta=-10^{\circ}$ and $\theta=39^{\circ}$ for the $(\eta,\,\eta')$ 
system and for $(\omega,\,\phi)$ system, respectively, while for other 
nonets we used $\theta=28^{\circ}$.

\medskip

For each experiment, the most recent measurements have been considered. 
Multiple measurements from different experiments have been averaged 
according to the PDG method \cite{pdg}, with error rescaling in case of 
discrepancy, that is a $\chi^2/dof$ of the weighted average $>1$. The 
overall calculated yields $T_i$ are compared to the experimental measurements 
$E_i$, and the total overall $\chi^2$, 
\begin{equation} 
  \chi^2 = \sum_i (T_i-E_i)^2/\sigma_i^2 
\end{equation}
where $\sigma_i$ are the experimental errors, is minimized. The minimization 
is in fact carried out in two steps, in order to take into account the 
uncertainties on input parameters, such as hadron masses, widths and branching 
ratios, according to the following procedure \cite{Beca-h}.  
First a $\chi^2$ with only experimental errors is minimised and preliminary 
best-fit model parameters are determined. 
Then, keeping the preliminarly fitted parameters fixed, the variations 
$\Delta n_j^{l \, {\rm theo}}$ of the multiplicities corresponding to 
the variations of the $l^{\rm th}$ input parameter 
by one standard deviation are calculated. Such variations are considered as 
additional systematic uncertainties on the multiplicities and the following 
covariance matrix is formed,
\begin{equation}
   {\sf C}_{ij}^{\rm sys} = \sum_l \Delta n_i^l  \Delta n_j^l ,
\end{equation}
to be added to the experimental covariance matrix ${\sf C}^{\rm exp}$. 
Finally a new $\chi^2$ is minimised with covariance matrix 
$\sf C^{\rm exp}+{\sf C}^{\rm sys}$, from which the best-fit estimates of the 
parameters and their errors are obtained. Actually, more than 350 among the 
most relevant or poorly known input parameters have been varied. However,
it should be mentioned that no variation of the branching fractions of 
heavy flavoured hadrons has been done. Therefore, for some specific species,
the systematic error could have been underestimated.

\section{Results}

\subsection{Light flavoured hadrons}

We begin our analysis with the most extensive sample, the LEP data at 
91.25 GeV. It comes from a compilation of results from the 
four different experimental groups (see references at the end of the
Appendix), and it lists up to 30 different light flavoured 
species. However, for short-lived and hence broad resonances, the separation 
of resonance signal from background often becomes difficult, making the 
assessment of systematic errors problematic. Moreover, broad resonances
yields are more sensitive to feeding from possibly unobserved heavier 
states or poorly known branching ratios. For this reason, we first 
consider, both for the conventional and for the Hawking-Unruh scenario,
the analysis of the unproblematic (``golden'') species of widths less than 
10 MeV; this still leaves 15 different hadronic states to be analysed,
listed in tables~\ref{91gold} and \ref{91bhgold}.
\begin{figure}[h]
\begin{minipage}[t]{7.5cm}
\epsfig{file=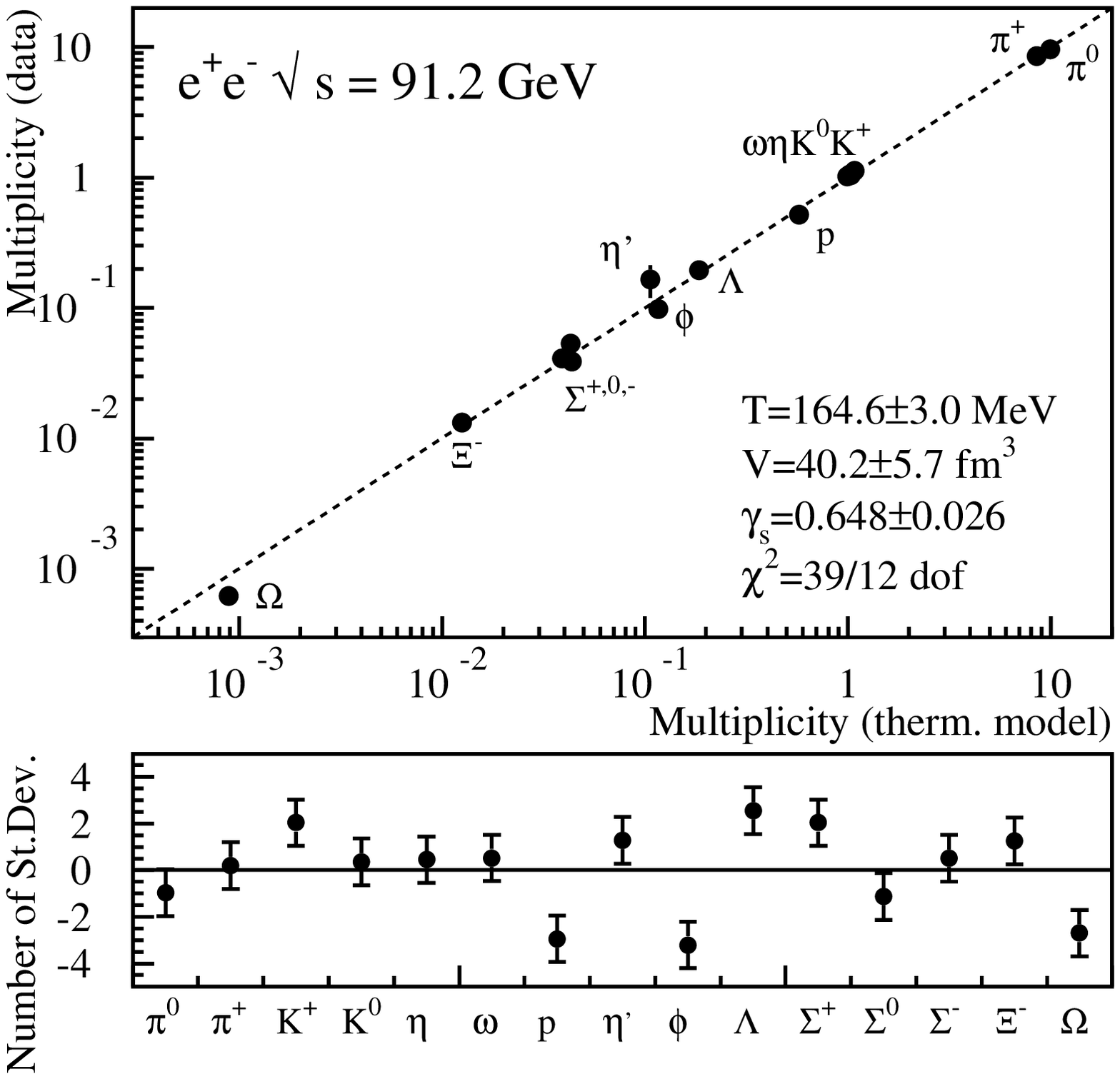,width=7.5cm}
\end{minipage}
\begin{minipage}[t]{7.5cm}
\epsfig{file=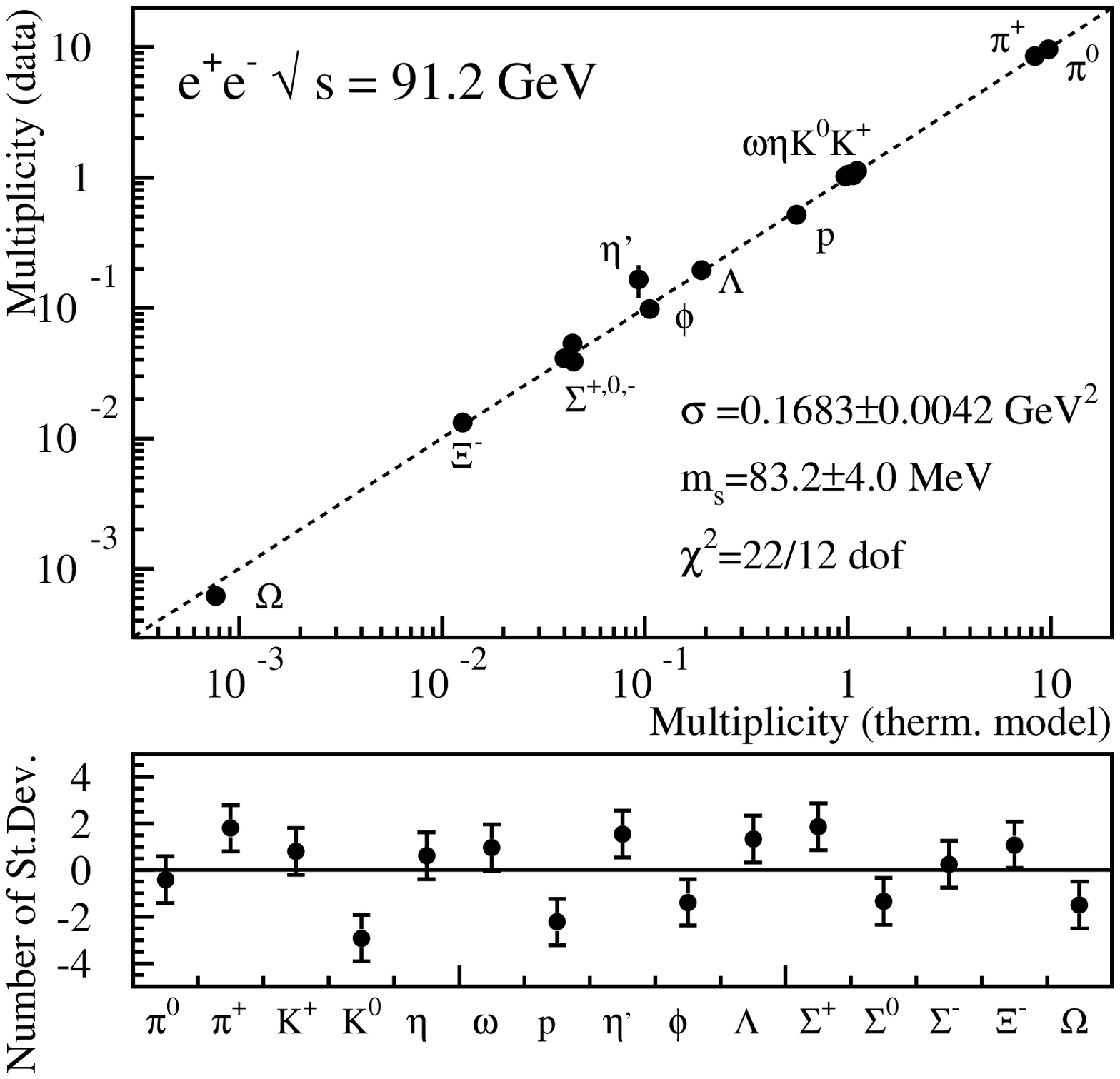,width=7.5cm}
\end{minipage}
\caption{Comparison between measured and fitted multiplicities of long-lived
hadronic species in \epe~collisions at $\sqrt s = 91.25$ GeV. Left: 
statistical hadronization model with one temperature. Right: Hawking-Unruh
radiation model.}
\label{ee91plot}
\end{figure}

\medskip

As noted, the conventional statistical resonance gas approach is based 
on a universal temperature $T$, a strangeness suppression factor $\gamma_s$, 
and a global volume $V$. The fit of the long-lived species is shown in 
detail in table~\ref{91gold} and fig.~\ref{ee91plot} and the resulting 
fit parameters are
\be\label{conv}
T = 164.6 \pm 3.0 ~{\rm MeV};~~ \gamma_s = 0.648 \pm 0.026;
~~V= 40.2 \pm 5.7~{\rm fm}^3,
\ee
with a $\chi^2$/dof = 39/12. The errors on the parameters are the fit 
errors rescaled by $\sqrt{\chi^2/dof}$. Such a method \cite{pdg} 
takes into account the additional uncertainty on the parameters if the
fit leads to $\chi^2/dof > 1$. This rescaling has been applied
to all parameter errors quoted in this paper.

\medskip

Next, we perform the corresponding hadron-resonance gas analysis in the
Hawking-Unruh formulation, introducing different temperatures determined 
by the string tension $\sigma$ and the strange quark mass $m_s$. The 
results for long-lived species are shown in table~\ref{91bhgold}
and fig.~\ref{ee91plot}. The resulting fit parameters here are
\be\label{hueq}
\sigma = 0.1683 \pm  0.0048~{\rm GeV}^2;~~m_s = 0.083 \pm 0.004~{\rm GeV},~~
V= 40.3 \pm 3.2~{\rm fm}^3;~~
\ee
with a  $\chi^2$/dof = 22/12, somewhat better than that of the corresponding 
conventional fit. 

\medskip

We now repeat both analyses using the entire 91.25 GeV data set, with the
results shown in table XX and XXI of the Appendix. The resulting fit values 
(see tables~\ref{91gold} and \ref{91bhgold}) agree well within errors with 
those obtained from the ``golden'' data  set at 91.25 GeV. As expected, because 
of the mentioned error sizes, the $\chi^2/dof$ for the full 91.25 set is 
considerably worse.
 
\begin{table}[h]
\begin{center}
\begin{tabular}{|c|c|c|c|c|}
\hline
                   & Experiment (E)           & Model (M)    & Residual        & $|$M - E$|$/E  [\%] \\ 
\hline
$\pi^0$            &	 9.61 \tpm	0.29  &        9.89  &        0.97     &       2.95    \\ 
$\pi^+$            &	 8.50 \tpm	0.10  &        8.48  &       -0.14     &     -0.167    \\ 
$K^+$              &	1.127 \tpm     0.026  &       1.074  &        -2.0     &      -4.69    \\ 
$K_S^0$            &   1.0376 \tpm    0.0096  &      1.0342  &       -0.35     &     -0.327    \\ 
$\eta$             &	1.059 \tpm     0.086  &       1.020  &       -0.46     &      -3.72    \\ 
$\omega$           &	1.024 \tpm     0.059  &       0.993  &       -0.52     &      -2.99    \\ 
$p$                &	0.519 \tpm     0.018  &       0.572  &         3.0     &       10.3    \\ 
$\eta'$            &	0.166 \tpm     0.047  &       0.106  &        -1.3     &      -36.4    \\ 
$\phi$             &    0.0977 \tpm   0.0058  &      0.1163  & 	     3.2       &	19.0	\\ 
$\Lambda$          &   0.1943 \tpm    0.0038  &      0.1846  &        -2.5     &      -4.98    \\ 
$\Sigma^{+}$       &  0.0535 \tpm    0.0052   &      0.0429  &	     -2.0     &      -19.9    \\ 
$\Sigma^{0}$       &  0.0389 \tpm    0.0041   &      0.0435  &	      1.1     &       11.8    \\ 
$\Sigma^{-}$       &  0.0410 \tpm    0.0037   &      0.0391  &	    -0.51     &      -4.58    \\ 
$\Xi^-$            & 0.01319 \tpm   0.00050   &     0.01256  &	     -1.3     &      -4.81    \\ 
$\Omega$           & 0.00062 \tpm   0.00010   &     0.00089  &	      2.7     &       43.7    \\ 
\hline
\end{tabular}
\caption{Abundances of long-lived hadrons in \epe~collisions at 
$\sqrt s$ = 91.25 GeV, compared to a statistical hadronization model fit
based on $T$ and $\gamma_s$. The third column shows the residual, defined
as the difference between model and data divided by the error, while fourth 
column shows the differences between model and data in percent.
References to the original experimental publications can be found in table
~\ref{ee91} in the Appendix.}
\label{91gold}
\end{center}
\end{table}

\begin{table}[h]
\begin{center}
\begin{tabular}{|c|c|c|c|c|}
\hline
                   & Experiment (E)            & Model (M)    &  Residual       &  $|$M - E$|$/E  [\%]\\ 
\hline
$\pi^0$            &      9.61 \tpm      0.29  &        9.73  &        0.41     &       1.25    \\ 
$\pi^+$            &      8.50 \tpm      0.10  &        8.32  &        -1.7     &      -2.06    \\ 
$K^+$              &     1.127 \tpm     0.026  &       1.106  &       -0.80     &      -1.85    \\ 
$K_S^0$            &    1.0376 \tpm    0.0096  &      1.0656  &         2.9     &       2.69    \\ 
$\eta$             &     1.059 \tpm     0.086  &       1.006  &       -0.61     &      -4.98    \\ 
$\omega$           &     1.024 \tpm     0.059  &       0.967  &       -0.97     &      -5.58    \\ 
$p$                &     0.519 \tpm     0.018  &       0.559  &         2.2     &       7.78    \\ 
$\eta'$            &     0.166 \tpm     0.047  &       0.093  &        -1.6     &      -43.8    \\ 
$\phi$             &    0.0977 \tpm    0.0058  &      0.1057  &         1.4     &       8.11    \\ 
$\Lambda$          &    0.1943 \tpm    0.0038  &      0.1892  &        -1.3     &      -2.63    \\ 
$\Sigma^{+}$       &    0.0535 \tpm    0.0052  &      0.0438  &        -1.9     &      -18.2    \\ 
$\Sigma^{0}$       &    0.0389 \tpm    0.0041  &      0.0444  &         1.4     &       14.2    \\ 
$\Sigma^{-}$       &    0.0410 \tpm    0.0037  &      0.0401  &       -0.25     &      -2.22    \\ 
$\Xi^-$            &   0.01319 \tpm   0.00050  &     0.01265  &        -1.1     &      -4.11    \\ 
$\Omega$           &   0.00062 \tpm   0.00010  &     0.00077  &         1.5     &       23.5    \\ 
\hline
\hline
\end{tabular}
\caption{Abundances of long-lived hadrons in \epe~collisions at 
$\sqrt s$ = 91.25 GeV, compared to a Hawking-Unruh fit in terms of
string tension $\sigma$ and strange quark mass $m_s$. The third column 
shows the residual, defined
as the difference between model and data divided by the error, while fourth 
column shows the differences between model and data in percent.
References to the original experimental publications can be found in table
~\ref{ee91bh} in the Appendix}
\label{91bhgold}
\end{center}
\end{table}

Here a comment is in order. The simple formulae (\ref{mult}) and (\ref{hfmult}),
in both models, rely on some side assumptions (e.g. the special distributions 
for cluster charge fluctuations needed for the introduction of the equivalent
global cluster) that are not expected to be exactly fulfilled. Therefore, those 
formulae are to be taken as a zero-order approximation and not as a faithful 
representation of the real process. Deviations from the introduced assumption
entail corrections to the formulae (\ref{mult}) and (\ref{hfmult}) which are 
nevertheless very difficult to estimate. The theoretical error involved in these 
formulae becomes important when the accuracy of measurements is comparable 
and, in this case, a bad $\chi^2$ is to be expected. This is probably
the case at $\sqrt s = 91.25$ GeV, where the relative accuracy of measurements 
is of the order of few percent for many particles. In this case, the $\chi^2$
fit is a useful tool to determine the best parameters of the ``simplified" theory
but should be used very carefully as a measure of the fit quality. As has been
mentioned, in order to take into account the uncertainty on parameters implied 
in fits with $\chi^2/dof>1$, parameter errors have been rescaled by $\chi^2/dof$ 
if this is larger than 1, according to Particle Data Group procedure \cite{pdg}.
\medskip

For all the remaining energies we have also carried out the corresponding 
analyses; the results are listed in tables~\ref{parameters1} and \ref{parameters2}
for the model parameters, while the comparison between measured and calculated
multiplicities are shown in tables X to XXXI of the Appendix.  

\begin{table}[!h]
\begin{center}
\begin{tabular}{|c|c|c|c|c|}
\hline
\sqrtsnn &  $T$ [MeV]             &  $VT^3$            & $\gamma_S$          & $\chi^2$/dof   \\
\hline
14       &       172.1 \tpm  5.2  &   8.3  \tpm   1.0  &  0.772 \tpm  0.094  &  0.9 /  3    \\
22       &       178.7 \tpm  3.7  &   8.70 \tpm   0.94 &  0.76  \tpm  0.10   &  0.7 /  3 \\
29       &       164.0 \tpm  5.4  &   15.0 \tpm   2.4  &  0.683 \tpm  0.075  &  33 / 13   \\
35       &       163.3 \tpm  3.2  &   15.0 \tpm   1.4  &  0.730 \tpm  0.045  &  8.2 /  7   \\
43       &       169   \tpm   10  &   13.5 \tpm   3.2  &  0.741 \tpm  0.074  & 2.9 /  3   \\
91       &       161.9 \tpm  4.1  &   25.8 \tpm   3.4  &  0.638 \tpm  0.039  & 215 / 27   \\
91*      &       164.6 \tpm  3.0  &   23.3 \tpm   2.2  &  0.648 \tpm  0.026  &  39 / 12   \\
133      &       167.1 \tpm  7.5  &   26.0 \tpm   4.6  &  0.671 \tpm  0.074  &  0.1 /  2   \\
161      &       153.4 \tpm  6.5  &   37.2 \tpm   5.9  &  0.72  \tpm  0.12   &  0.03 /  1   \\
183      &       161 \tpm    13   &   35   \tpm   11   &  0.446 \tpm  0.098  &  5.0 /  2    \\
189      &       159 \tpm    12   &   36   \tpm   10   &  0.54  \tpm  0.11   &  7.5 /  2   \\
\hline
\end{tabular}
\caption{Best fit parameters for the statistical hadronization model in 
\epe~collisions. The golden sample fit is marked with a $*$.}
\label{parameters1}
\end{center}
\end{table}
\begin{table}[!h]
\begin{center}
\begin{tabular}{|c|c|c|c|c|}
\hline
\sqrtsnn & $\sigma$ [GeV$^2$]       &$V\sigma^{3/2}$   &$m_s$ [MeV]        & $\chi^2$/dof \\
\hline
14       &      0.185 \tpm  0.015   &  133 \tpm   24   &    71   \tpm 19   &   0.9 /  3   \\
22       &      0.199 \tpm  0.017   &  140 \tpm   32   &    77   \tpm 20   &   0.7 /  3   \\
29       &      0.1673 \tpm 0.0096  &  240 \tpm   36   &    78   \tpm 15   &     38 / 13   \\
35       &      0.1675 \tpm 0.0065  &  237 \tpm   23   &    74.6 \tpm 6.5  &   8.8 /  7	  \\
43       &      0.178  \tpm 0.020   &  216 \tpm   48   &    76   \tpm 16   &   3.2 /  3   \\
91       &      0.1625 \tpm 0.0078  &  406 \tpm   52   &    82.3 \tpm 7.8  &   217 / 27   \\
91*      &      0.1683 \tpm 0.0042  &  368 \tpm   24   &    83.2 \tpm 4.0  &   23 / 12    \\
133      &      0.175 \tpm  0.015   &  418 \tpm   69   &    89   \tpm 16   &   0.1 /  2   \\
161      &      0.148 \tpm  0.029   &  590 \tpm   220  &    74   \tpm 24   &   0.03 /  1  \\
183      &      0.165 \tpm  0.026   &  550 \tpm   160  &    130  \tpm 28   &   5.1 /  2   \\
189      &      0.161 \tpm  0.029   &  560 \tpm   180  &    110  \tpm 36   &   7.7 /  2   \\
\hline
\end{tabular}
\caption{Best fit parameters for the Hawking-Unruh model in 
\epe~collisions. The golden sample fit is marked with a $*$.}
\label{parameters2}
\end{center}
\end{table}

\medskip

\begin{figure}[h]
\begin{minipage}[t]{7.5cm}
\epsfig{file=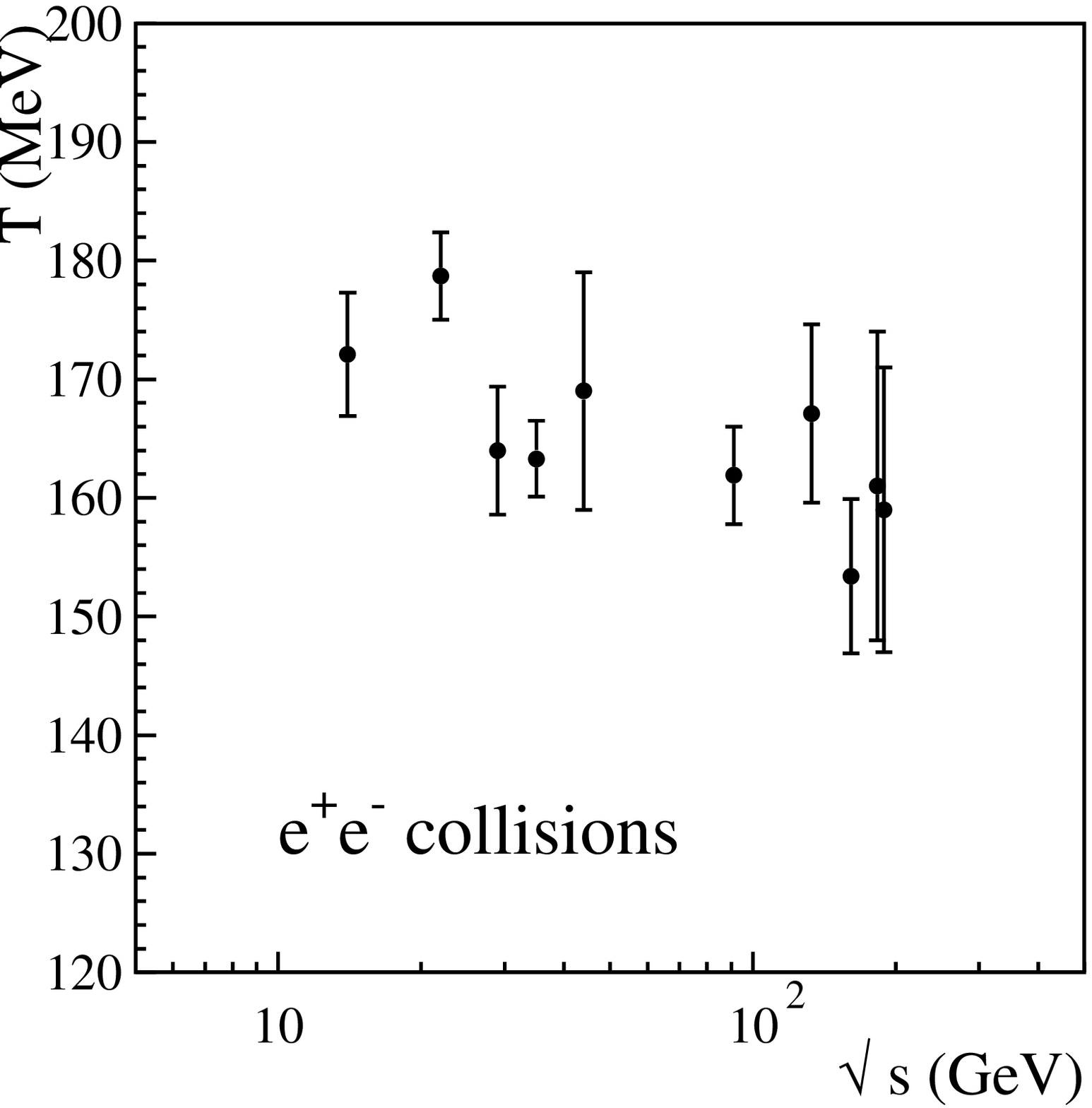,width=7.5cm}
\end{minipage}
\begin{minipage}[t]{7.5cm}
\epsfig{file=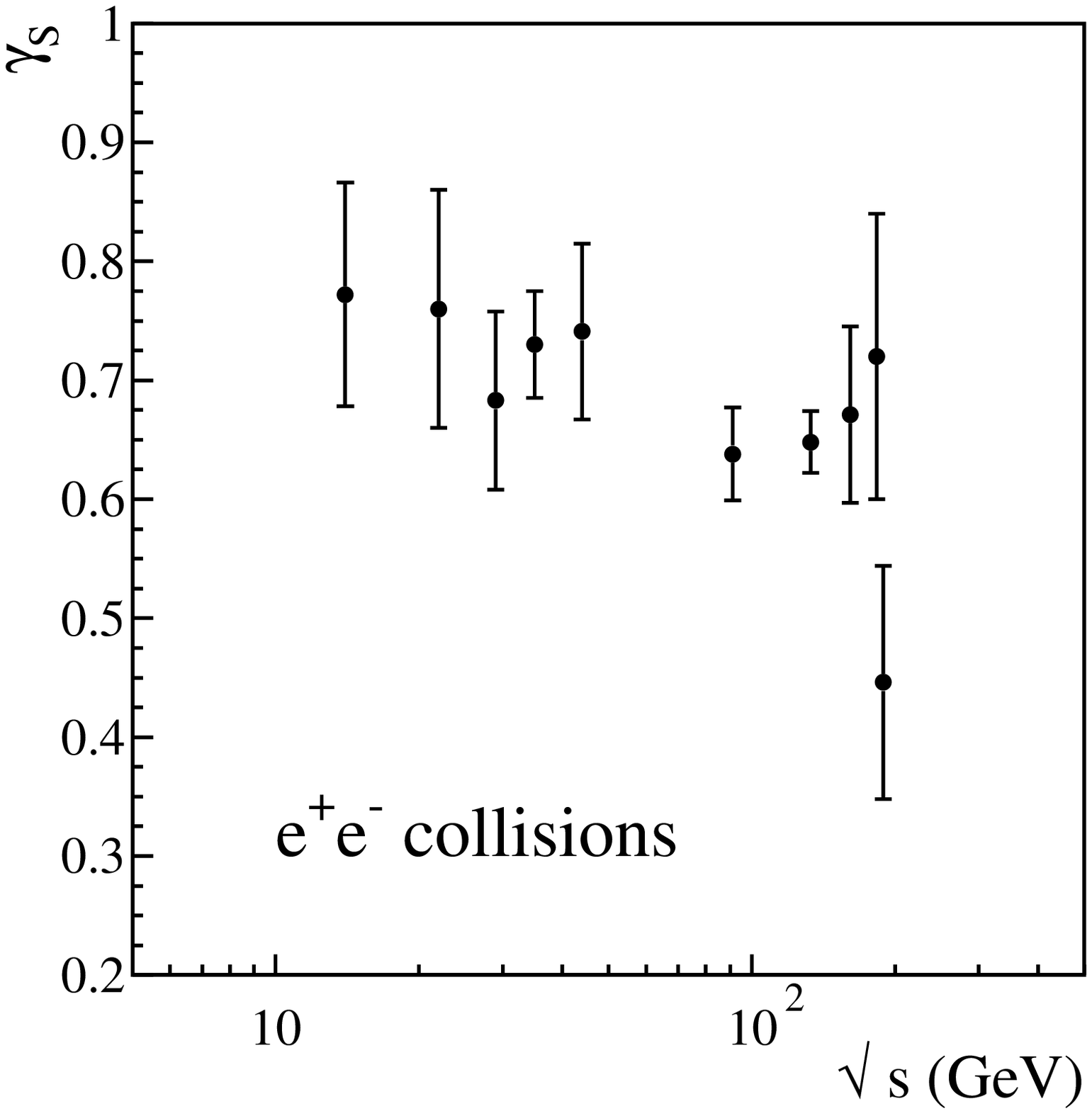,width=7.5cm}
\end{minipage}
\caption{Hadronization temperature $T$ (left) and strangeness suppression 
factor $\gamma_s$ (right) from conventional statistical fits to hadron 
abundances in $e^+e^-$ annihilation, as function of the incident energy 
$\sqrt s$. }
\label{temp/gam}
\end{figure}

\begin{figure}[h!]
\begin{minipage}[t]{7.5cm}
\epsfig{file=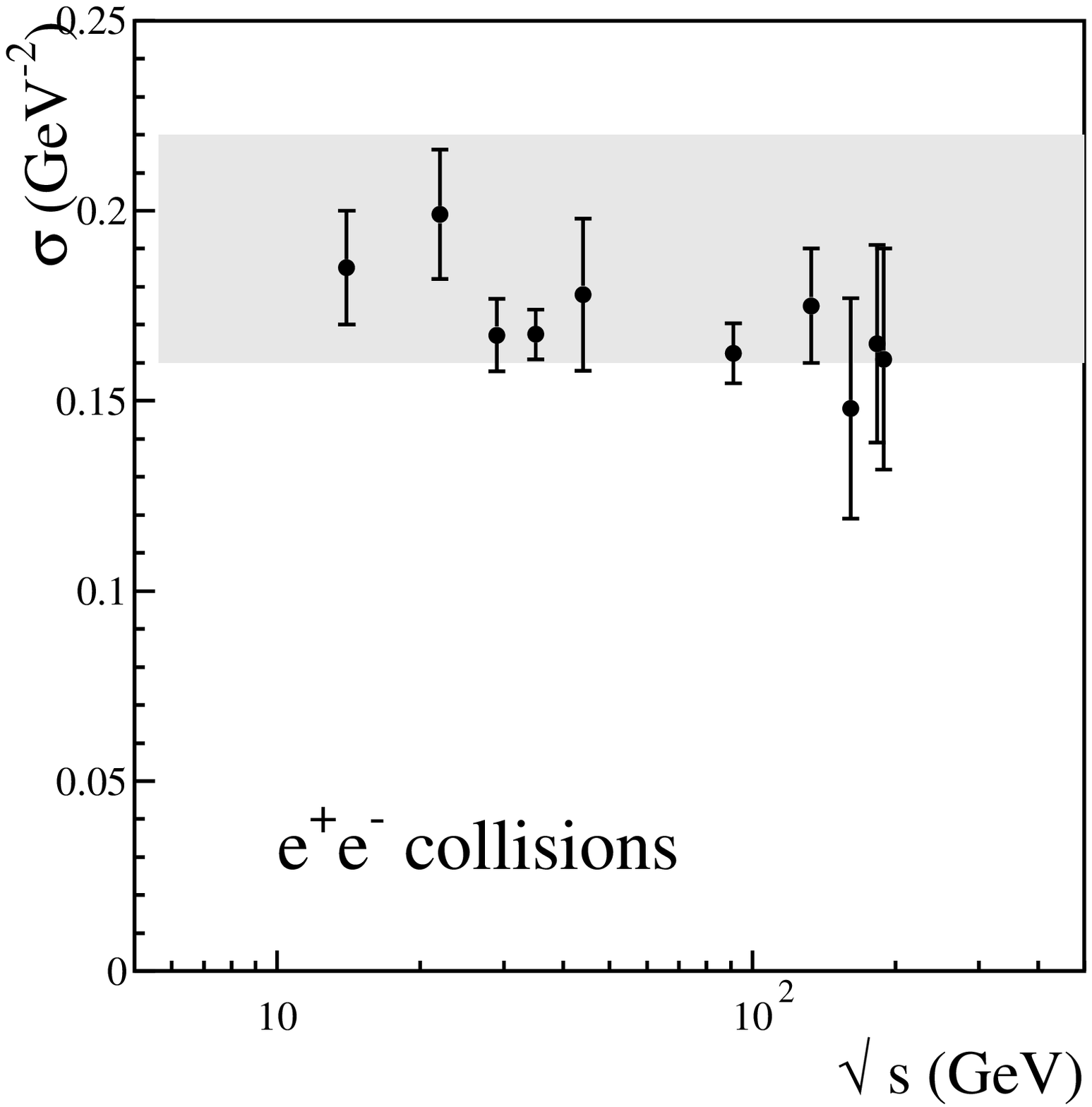,width=7.5cm}
\end{minipage}
\begin{minipage}[t]{7.5cm}
\epsfig{file=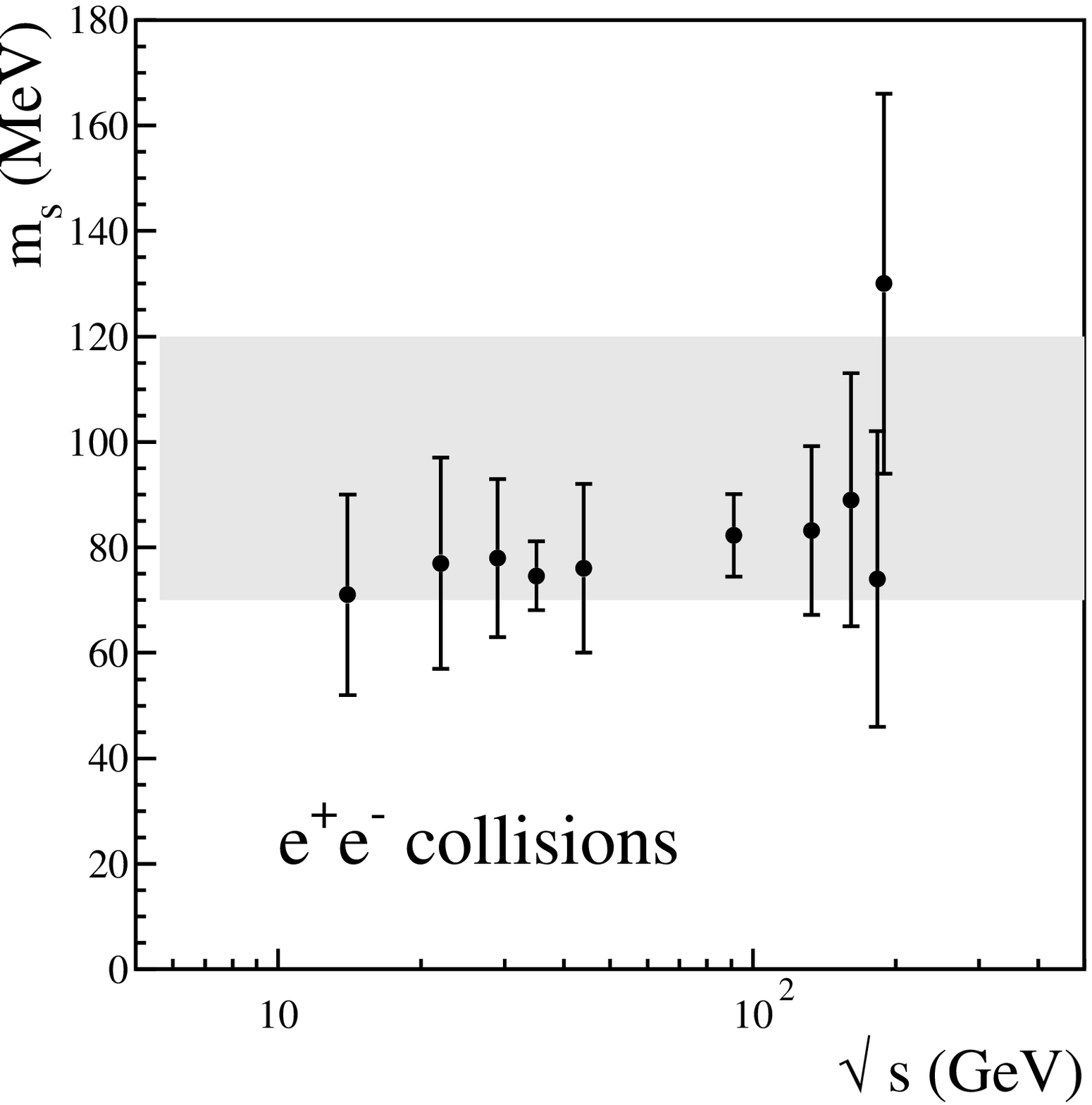,width=7.5cm}
\end{minipage}
\caption{String tension (left) and strange quark mass (right) from 
Hawking-Unruh fits to hadron abundances in $e^+e^-$ annihilation, 
as function of the incident energy $\sqrt s$. The shaded bands give
the overall average values as determined by other data.}
\label{string/mass}
\end{figure}

The parameters $T,\gs$ and $\sigma,m_s$ are shown in figs. \ref{temp/gam} and 
\ref{string/mass} respectively. It can be seen that they are remarkably 
constant throughout the examined energy range from $\sqrt{s}=14$ to 189 GeV. 


We now recall that the quantities we treated as fit parameters in the 
Hawking-Unruh analyses, the string tension and the strange quark mass, have 
in fact been determined in various other contexts and by different methods; 
they are quite well-known. The string tension $\sigma$ is obtained in studies 
of heavy quarkonium spectroscopy as well as from Regge phenomenology. The 
canonical value \cite{JOS} was $0.192$ GeV$^2$; more recent calculations 
range from $0.16$ GeV$^2$ \cite{Nora} 
to 0.22 GeV$^2$ \cite{BBC}, giving an estimate of $\sigma = 0.19 \pm 0.03$.
The best average value of the strange quark mass is presently listed as
\cite{pdg} $m_s= 0.095 \pm 0.025~{\rm GeV}$. In both cases we have good 
agreement with our fit values, as seen in fig.~\ref{string/mass}. We can
thus indeed conclude that the Hawking-Unruh approach provides 
a parameter-free description of thermal hadron abundances in $e^+e^-$ 
annihilation. The suppression of hadrons containing strange quarks
is fully accounted for in terms of slight temperature changes due to 
the heavier strange quark mass. It is thus natural that this affects 
also non-strange hadrons dominantly made up of a strange quarks, such 
as the $\phi$. 

\medskip

To illustrate the effect, we list in table~\ref{texample} the different temperatures 
resulting from the different strange quark contents of the observed hadrons, 
using the fit parameters from the golden 91.25 data set. To check what
such temperature differences can lead to, we compare the rate of direct $\phi$ 
production in the conventional to that of the Hawking-Unruh scenario. This direct 
rate is given by
\be
\langle n \rangle_\phi = 3 {V T m^2 \over 2 \pi^2} {\rm K}_2(m/T) 
 ~\gamma_S^2
\ee
in the conventional scenario; using the values (\ref{conv}) together with
the production value listed in table~\ref{parameters1}, we obtain
$\langle n \rangle_\phi \simeq 0.078$. Note that $\gamma_s^2\simeq 0.42$ 
reduces the equilibrium value by more than a factor of two.
The Hawking-Unruh scheme has with $T(ss)=0.148$
GeV a lower temperature for a meson containing a strange quark-antiquark
pair than that governing light quark mesons. With
\be
 \langle n \rangle_\phi = 3 {V T(ss) m^2 \over 2 \pi^2} {\rm K}_2(m/T(ss))
\ee
and the corresponding production volume of table~\ref{parameters2},
this leads to $\langle n \rangle_\phi \simeq 0.077$ and 
hence practically the same value, however, without invoking the parameter 
$\gamma_s$. We note that these results should not be compared directly to the
$\phi$ production measured in $e^+e^-$ annihilation, which contains
(at 91.25 GeV) a further 30 \% due to feed-down contributions from
charmed and bottomed hadron decay.  
\begin{table}[h]
\begin{center}
\begin{tabular}{|c|c|}  \hline
\hline
$ T  $ {\rm[GeV]}  &  \\
\hline
$T(00)$  &   0.164  \\
$T(0s)$  &   0.156  \\
$T(ss)$  &   0.148  \\
$T(000)$ &   0.164  \\
$T(00s)$ &   0.158  \\
$T(0ss)$ &   0.153  \\
$T(sss)$ &   0.148  \\
\hline
\end{tabular}
\caption{Hadronization temperatures for hadrons of different strangeness 
content, for $m_s= 0.083$ Gev and $\sigma=0.169$ GeV$^2$.}
\label{texample}
\end{center}
\end{table}

\subsection{Heavy flavoured hadrons}

As has been mentioned in the previous section, the calculation of 
heavy flavoured
hadron yields in \eecc~ and \eebb~ events is necessary to determine the final
light flavoured hadron multiplicities. The heavy flavoured hadron relative
abundances in such events are determined according to formula (\ref{hfmult}). 
For the conventional formulation of the statistical model, one has the same 
temperature as for the light-flavoured hadron species, while for the Hawking-
Unruh radiation model, we do not have a definite prescription and we chose
to use the same temperatures as for the light flavoured hadrons, which are
dependent on the quark content.

\medskip

Once the model parameters have been fitted by using light-flavoured hadronic
multiplicities, they can be used to predict relative yields of heavy flavoured 
species in \eecc~and \eebb~annihilations and compare them to measured ones. 
This is a powerful, parameter-free, independent test of the conventional
statistical model and a necessary consistency check for the Hawking-Unruh 
radiation model. Also, comparing theoretical values to measured relative 
abundances of heavy flavoured hadrons in specific annihilations channels 
(e.g. charmed hadrons in \eecc) we achieve a more effective test because, 
e.g., the contribution of weak $b \to c$ decays is excluded.

\medskip

The relative yields of several heavy flavoured hadronic species have been 
measured 
in \epe~collisions at 91.25 GeV by the four LEP experiments. We show a 
comparison between model and weighted averaged experimental values in tables 
\ref{hfstat} and \ref{hfhu}. 

\medskip

For the statistical model (table~\ref{hfstat}), the theoretical values have 
been estimated by using the parameters in eq.~(\ref{conv}). The agreement 
between model and experiment 
is strikingly good, with few peculiar deviations in heavier states. In all 
those cases we observe an underestimation of measured values which may partly 
be explained by the absence, in our input spectrum, of unknown heavier heavy 
flavoured resonances feeding these states. It is quite remarkable that 
the model is able to reproduce the largely different $V/P$ ratios in the charm ($D^*/D$) 
and bottomed ($B^*/B$) sector - a long-standing issue in string models - 
without any additional parameter. This result confirms previous early findings
\cite{Beca-e,santorini}.
 
\begin{table}[!h]
\begin{center}
\begin{tabular}{|c|c|c|c|c|c|}
\hline
   Particle  &     &   Experiment (E)   & Model (M)  & Residual & $(M - E)/E$  [\%] \\ 
\hline

$D^0$        & \cite{99bg}         & 0.559 \tpm 0.022	& 0.5406     & -0.83    & -3.2  	\\
$D^+$        & \cite{99bg}         & 0.238 \tpm 0.024   & 0.2235     & -0.60    & -6.1  	\\
$D^{*+}$     &\cite{97ki,99bg,99vx}& 0.2377\tpm 0.0098  & 0.2279     & -1.00    & -4.1  	\\
$D^{*0}$     & \cite{98as}         & 0.218 \tpm  0.071	& 0.2311     & 0.18     &  6.0  	\\
$D^0_1$      & \cite{alconf,97vc}  & 0.0173 \tpm 0.0039 & 0.01830    & 0.26     &  5.8  	\\
$D^{*0}_2$   & \cite{alconf,97vc}  & 0.0484 \tpm 0.0080 & 0.02489    & -2.94    & -48.6 	\\
$D_s$        & \cite{99bg}         & 0.116  \tpm 0.036  & 0.1162     &  0.006   &  0.19 	\\
$D^*_s$      & \cite{99bg}         & 0.069  \tpm 0.026  & 0.0674     & -0.06    & -2.4  	\\
$D_{s1}$     & \cite{01nj,97vc}    & 0.0106 \tpm 0.0025 & 0.00575    & -1.94    & -45.7 	\\
$D^*_{s2}$   & \cite{01nj}         & 0.0140 \tpm 0.0062 & 0.00778    & -1.00    & -44.5 	\\
$\Lambda_c$  & \cite{99bg}         & 0.079  \tpm 0.022  & 0.0966     & 0.80     & 22.2  	\\
\hline \hline
$(B^0+B^+)/2$                &	\cite{Abbaneo:2001bv}      &  0.399 \tpm 0.011	&  0.3971    & -0.18   &  -0.49     \\
$B_s$                        &	\cite{Abbaneo:2001bv}      &  0.098 \tpm 0.012	&  0.1084    &  0.87   &  10.6	    \\
$B^*/B$(uds)                 &	\cite{96gz,95mt,95ky,94qv} &  0.749 \tpm 0.040	&  0.6943    & -1.37   &  -7.3	    \\
$B^{**}\times BR(B(^*)\pi)$  &	\cite{98cq,delconf,94fz}   &  0.180 \tpm 0.025  &  0.1319    & -1.92   &  -26.7     \\
$(B^*_2+B_1)\times BR(B(^*)\pi)$& \cite{delconf}           &  0.090 \tpm 0.018  &  0.0800    & -0.57   &  -11.4     \\
$B^*_{s2} \times BR(BK)$     &	 \cite{delconf}            &  0.0093 \tpm 0.0024&  0.00631   & -1.24   &  -32.1     \\
b-baryon                     &	\cite{Abbaneo:2001bv}      &  0.103 \tpm  0.018	&  0.09751   & -0.30   &  -5.3	    \\
$\Xi_b^-$                    &	\cite{Abbaneo:2001bv}      &  0.011 \tpm  0.006	&  0.00944   & -0.26   &  -14.2     \\

\hline	
\end{tabular}
\bigskip
\caption{Abundances of charmed hadrons in \eecc~annihilations and bottomed
hadrons in \eebb~annihilations at $\sqrt s$ = 91.25 GeV, compared to the 
prediction of the statistical model.}
\label{hfstat}
\end{center}
\end{table}
 
\medskip

For the Hawking-Unruh radiation model (table~\ref{hfhu}), the agreement between 
theoretical multiplicities, calculated with the parameters in eq.~(\ref{hueq}) and 
experiment is generally good, although not as good as for the conventional scheme. 
In fact, there are some specific discrepancies, especially in the beauty sector. 
In particular, we underestimate the relative yield of $B_s$ mesons, which is an 
effect of the lower temperature for open strange particles combined with the 
high mass of these particles, compared to light-flavoured species. In general, the 
way heavy flavoured hadrons are to be calculated in this model is still an open issue.

\begin{table}[!h]
\begin{center}
\begin{tabular}{|c|c|c|c|c|c|}
\hline
   Particle  &     &   Experiment (E)   & Model (M)  & Residual & $(M - E)/E$  [\%] \\ 
\hline

$D^0$        & \cite{99bg}         & 0.559 \tpm 0.022	&  0.5635   &  0.20   &  0.81  \\
$D^+$        & \cite{99bg}         & 0.238 \tpm 0.024   &  0.2332   & -0.20   &  -2.0  \\
$D^{*+}$     &\cite{97ki,99bg,99vx}& 0.2377\tpm 0.0098  &  0.2373   & -0.04   & -0.18  \\
$D^{*0}$     & \cite{98as}         & 0.218 \tpm  0.071	&  0.2407   &  0.32   &  10.4  \\
$D^0_1$      & \cite{alconf,97vc}  & 0.0173 \tpm 0.0039 &  0.01897  &  0.43   &   9.6  \\
$D^{*0}_2$   & \cite{alconf,97vc}  & 0.0484 \tpm 0.0080 &  0.02577  & -2.82   & -46.8  \\
$D_s$        & \cite{99bg}         & 0.116  \tpm 0.036  &  0.08460  & -0.87   & -27.1  \\
$D^*_s$      & \cite{99bg}         & 0.069  \tpm 0.026  &  0.04793  & -0.81   & -30.5  \\
$D_{s1}$     & \cite{01nj,97vc}    & 0.0106 \tpm 0.0025 &  0.00356  & -2.82   & -66.5  \\
$D^*_{s2}$   & \cite{01nj}         & 0.0140 \tpm 0.0062 &  0.00479  & -1.49   & -66.1  \\
$\Lambda_c$  & \cite{99bg}         & 0.079  \tpm 0.022  &  0.09922  &  0.92   &  25.6  \\
\hline\hline
$(B^0+B^+)/2$                &	\cite{Abbaneo:2001bv}      &  0.399 \tpm 0.011	&  0.4390   &  3.64 &  10.0     \\
$B_s$                        &	\cite{Abbaneo:2001bv}      &  0.098 \tpm 0.012	&  0.0276   & -5.87 & -71.9    \\
$B^*/B$(uds)                 &	\cite{96gz,95mt,95ky,94qv} &  0.749 \tpm 0.040	&  0.6978   & -1.28 & -6.8     \\
$B^{**}\times BR(B(^*)\pi)$  &	\cite{98cq,delconf,94fz}   &  0.180 \tpm 0.025  &  0.1479   & -1.28 & -17.8    \\
$(B^*_2+B_1)\times BR(B(^*)\pi)$& \cite{delconf}           &  0.090 \tpm 0.018  &  0.0894   & -0.04 & -0.72    \\
$B^*_{s2} \times BR(BK)$     &	 \cite{delconf}            &  0.0093 \tpm 0.0024&  0.00136  & -3.31 & -85.3    \\
b-baryon                     &	\cite{Abbaneo:2001bv}      &  0.103 \tpm  0.018	&  0.0944   & -0.48 & -8.4     \\
$\Xi_b^-$                    &	\cite{Abbaneo:2001bv}      &  0.011 \tpm  0.006	&  0.00415  & -1.14 & -62.2    \\

\hline
\end{tabular}
\bigskip
\caption{Abundances of charmed hadrons in \eecc~annihilations and bottomed hadrons 
in \eebb~annihilations at $\sqrt s$ = 91.25 GeV, compared to the prediction of 
the Hawking-Unruh radiation model.} 
\label{hfhu}										    
\end{center}
\end{table}

\section{Conclusions}

We have shown that in accord with previous studies \cite{Beca-e,erice,Beca-h},
the thermal hadron abundances observed in \epe~collisions over a wide range 
of energies can indeed be accounted for in an ideal resonance gas scenario, 
based on a universal temperature $T\simeq 165$ MeV and a strangeness suppression 
factor $\gamma_s \simeq 0.7$.
The latter is the {\sl ad hoc} price paid in order to account for the 
deviation from full chemical equilibrium observed in the data. Remarkably,
also the relative abundances of heavy flavoured species are in very
good agreement with the statistical-thermal {\em ansatz}.

\medskip

The Hawking-Unruh scenario, on the other hand, provides an intrinsic
deviation from full equilibrium through the dependence of the 
radiation temperature on the mass of the emitted quark. Given the value 
of this mass and the string tension specifying the field strength at the 
confinement horizon, we then have a parameter-free prediction of the
relative hadron abundances. We have seen here that these predictions
agree well with the data at all energies, with the {\sl caveat} that 
the relative multiplicities of heavy flavoured hadron species are not 
completely understood in this picture and are in slightly worse agreement
with respect to the conventional statistical model. In a subsequent paper, we 
shall extend the Hawking-Unruh description to high energy heavy ion collisions,
where it becomes significantly modified.

\medskip

In closing, we comment on the degree of agreement between experiment and
theory in our description. Hawking-Unruh radiation is thermal in leading
order \cite{P-W}, with higher order interaction terms. Similarly, one
expects corrections to the simplest statistical hadronization model
formulae, see discussion in Sect.~4. When accuracy of measurements is
good enough, such higher-order effects must be taken into account and the 
fit quality to the simplest formulae unavoidably degrades. It is 
an interesting question to see if nuclear collisions, with a higher degree 
of averaging, lead to smaller deviations with measurements of the same
accuracy.

\subsection*{Note added in proof}

After completion of this work, another analysis of hadron production in
$e^+e^-$ annihilation has appeared, reaching very different conclusions
\cite{pbmee}. However, in contrast to our work, it is assumed in that analysis
that the conservation of charm and bottom can be neglected.
In $e^+e^-$ annihilation, more than 40\% of all events contain a
primary charm or bottom quark-antiquark pair (see our table~\ref{bratios}), 
hence two heavy-flavoured hadrons. Neglecting the corresponding conservation 
conditions and the decay contributions of these heavy-flavoured hadrons
into light-flavoured hadrons necessarily produces serious disagreement with 
the data, especially for strange particles.

\clearpage
\section*{Appendix}
In the following Tables \ref{ee14} - \ref{ee189bh}, the experimental data
and the statistical hadronization model predictions for various hadron 
multiplicities are compared. The first column shows the experimental value
while in the second column, the statistical hadronization model prediction
is quoted. In the fourth column one can find the relative deviation of the 
model from the experiment in percentage while the third column shows the 
residuals defined as
\be
\mathrm{Residual}_i=\frac{N^{th}_i-N^{ex}_i}{\sigma_i}
\ee
in which $N^{th}_i$ and $N^{ex}_i$ are the theoretical and experimental multiplicitites
and $\sigma_i$ is the (experimental) standard error of a particle species $i$. For each of the 
energies, we show 2 different tables one after the other so that always the first table shows
the conventional statistical hadronization model fit results while the following one shows 
the same information in the Hawking-Unruh approach.
\bigskip

\begin{table}[!h]\begin{center}
\begin{tabular}{|c|c|c|c|c|c|}
\hline
        &        & Experiment (E)       & Model (M)  & Residual & (M - E)/E  (\%)\\ 
\hline
$\pi^0$            &\cite{Bartel:1985wn} &      4.69 \tpm      0.20  &        4.65  &       -0.18   &     -0.752    \\ 
$\pi^+$            &\cite{Althoff:1984iz} &      3.60 \tpm      0.30  &        3.79  &        0.62   &       5.18    \\ 
$K^+$              &\cite{Althoff:1984iz} &     0.600 \tpm     0.070  &       0.589  &       -0.15   &      -1.81    \\ 
$K_S^0$            &\cite{Althoff:1984iz,Bartel:1983qp} &     0.563 \tpm     0.045  &       0.556  &       -0.15   &      -1.17    \\ 
$p$                &\cite{Althoff:1984iz} &     0.210 \tpm     0.030  &       0.199  &       -0.37   &      -5.36    \\ 
$\Lambda$          &\cite{Althoff:1984iz} &     0.065 \tpm     0.020  &       0.077  &        0.58   &       17.8    \\ 
\hline
\end{tabular}\end{center}
\caption{
Hadron multiplicities in \epe~collisions at 14 GeV, compared to the outcomes of the fit
based on the statistical hadronization model with 
$T$ and $\gamma_S$. The third and fourth columns show the differences between 
data and model in units of standard error and in percentages, respectively.}\label{ee14}
\end{table}

\begin{table}[!h]\begin{center}
\begin{tabular}{|c|c|c|c|c|c|}
\hline
 &               & Experiment (E)       & Model (M)  & Residual & (M - E)/E  (\%)\\ 
\hline
$\pi^0$            &~\cite{Bartel:1985wn} &      4.69 \tpm      0.20  &        4.66  &       -0.17   &     -0.712    \\ 
$\pi^+$            &~\cite{Althoff:1984iz} &      3.60 \tpm      0.30  &        3.79  &        0.62   &       5.16    \\ 
$K^+$              &~\cite{Althoff:1984iz} &     0.600 \tpm     0.070  &       0.589  &       -0.16   &      -1.85    \\ 
$K_S^0$            &~\cite{Althoff:1984iz,Bartel:1983qp} &     0.563 \tpm     0.045  &       0.556  &       -0.14   &      -1.14    \\ 
$p$                &~\cite{Althoff:1984iz} &     0.210 \tpm     0.030  &       0.198  &       -0.39   &      -5.52    \\ 
$\Lambda$          &~\cite{Althoff:1984iz} &     0.065 \tpm     0.020  &       0.077  &        0.58   &       17.8    \\ 
\hline
\end{tabular}\end{center}
\caption{Hadron multiplicities in \epe~collisions at 14 GeV, 
compared to the outcomes of the fit based on Hawking-Unruh radiation model.
The third and fourth columns show the differences between 
data and model in units of standard error and in percentages, respectively.}\label{ee14bh}
\end{table}

\begin{table}[!h]\begin{center}
\begin{tabular}{|c|c|c|c|c|c|}
\hline
 &               & Experiment (E)       & Model (M)  & Residual & (M - E)/E  (\%)\\ 
\hline
$\pi^0$            &~\cite{Bartel:1985wn} &      5.50 \tpm      0.40  &        5.49  &      -0.033   &     -0.238    \\ 
$\pi^+$            &~\cite{Althoff:1984iz} &      4.40 \tpm      0.50  &        4.54  &        0.28   &       3.16    \\ 
$K^+$              &~\cite{Althoff:1984iz} &      0.75 \tpm      0.10  &        0.69  &       -0.65   &      -8.64    \\ 
$K_S^0$            &~\cite{Bartel:1983qp,Braunschweig:1989wg} &     0.638 \tpm     0.057  &       0.651  &        0.22   &       1.96    \\ 
$p$                &~\cite{Althoff:1984iz} &     0.310 \tpm     0.030  &       0.305  &       -0.16   &      -1.51    \\ 
$\Lambda$          &~\cite{Althoff:1984iz} &     0.110 \tpm     0.025  &       0.117  &        0.29   &       6.70    \\ 
\hline
\end{tabular}\end{center}
\caption{Hadron multiplicities in \epe~collisions at 22 GeV, compared to the outcomes of the fit
based on the statistical hadronization model with 
$T$ and $\gamma_S$. The third and fourth columns show the differences between 
data and model in units of standard error and in percentages, respectively.}\label{ee22}
\end{table}

\begin{table}[!h]\begin{center}
\begin{tabular}{|c|c|c|c|c|c|}
\hline
 &               & Experiment (E)       & Model (M)  & Residual & (M - E)/E  (\%)\\ 
\hline
$\pi^0$            &~\cite{Bartel:1985wn}  &      5.50 \tpm      0.40  &        5.49  &      -0.020   &     -0.148    \\ 
$\pi^+$            &~\cite{Althoff:1984iz} &      4.40 \tpm      0.50  &        4.54  &        0.28   &       3.18    \\ 
$K^+$              &~\cite{Althoff:1984iz} &      0.75 \tpm      0.10  &        0.68  &       -0.65   &      -8.71    \\ 
$K_S^0$            &~\cite{Bartel:1983qp,Braunschweig:1989wg} &     0.638 \tpm     0.057  &       0.651  &        0.22   &       1.94    \\ 
$p$                &~\cite{Althoff:1984iz} &     0.310 \tpm     0.030  &       0.305  &       -0.17   &      -1.65    \\ 
$\Lambda$          &~\cite{Althoff:1984iz} &     0.110 \tpm     0.025  &       0.118  &        0.31   &       7.02    \\ 
\hline
\end{tabular}\end{center}
\caption{Hadron multiplicities in \epe~collisions at 22 GeV,compared to the outcomes of the fit based on Hawking-Unruh radiation model. 
The third and fourth columns show the differences between 
data and model in units of standard error and in percentages, respectively.}\label{ee22bh}
\end{table}

\begin{table}[!h]\begin{center}
\begin{tabular}{|c|c|c|c|c|c|}
\hline
&                & Experiment (E)       & Model (M)  & Residual & (M - E)/E  (\%)\\ 
\hline
$\pi^0$            &\cite{Aihara:1984mi} &      5.30 \tpm      0.70  &        6.48  &         1.7   &       22.2    \\ 
$\pi^+$            &\cite{Aihara:1986mv} &      5.35 \tpm      0.25  &        5.42  &        0.26   &       1.23    \\ 
$K^+$              &\cite{Aihara:1986mv} &     0.700 \tpm     0.050  &       0.747  &        0.93   &       6.66    \\ 
$K_S^0$            &\cite{Schellman:1984yz,Aihara:1984mk,Abachi:1989pr} &     0.691 \tpm     0.029  &       0.712  &        0.73   &       3.05    \\ 
$\eta$             &\cite{Abachi:1987qd,Wormser:1988ru} &     0.584 \tpm     0.075  &       0.654  &        0.92   &       11.8    \\ 
$\rho^0$           &\cite{Abachi:1989em} &     0.900 \tpm     0.050  &       0.745  &        -3.1   &      -17.2    \\ 
$K^{*+}$           &\cite{Abachi:1987wc} &     0.310 \tpm     0.030  &       0.237  &        -2.4   &      -23.7    \\ 
$K^{*0}$           &\cite{Abachi:1989em,Aihara:1984mk} &     0.281 \tpm     0.022  &       0.232  &        -2.2   &      -17.3    \\ 
$p$                &\cite{Aihara:1986mv} &     0.300 \tpm     0.050  &       0.300  &      0.0038   &     0.0627    \\ 
$\eta'$            &\cite{Wormser:1988ru} &      0.26 \tpm      0.10  &        0.07  &        -1.8   &      -71.6    \\ 
$\phi$             &\cite{Aihara:1984pw} &     0.084 \tpm     0.022  &       0.092  &        0.35   &       9.15    \\ 
$\Lambda$          &\cite{Aihara:1984wx,delaVaissiere:1984xg,Geld:1992si} &    0.0983 \tpm    0.0060  &      0.1016  &        0.56   &       3.43    \\ 
$\Xi^-$            &\cite{Klein:1986ws,Abachi:1987ac} &    0.0083 \tpm    0.0020  &      0.0070  &       -0.64   &      -15.4    \\ 
$\Sigma^{*+}$      &\cite{Abachi:1987ac} &    0.0083 \tpm    0.0024  &      0.0111  &         1.2   &       34.2    \\ 
$K_2^{*+}$         &\cite{Abachi:1987wc} &     0.045 \tpm     0.022  &       0.016  &        -1.3   &      -64.4    \\ 
$\Omega$           &\cite{Klein:1987fu} &    0.0070 \tpm    0.0036  &      0.0005  &        -1.8   &      -93.4    \\ 
\hline
\end{tabular}\end{center}
\caption{Hadron multiplicities in \epe~collisions at 29 GeV, compared to the outcomes of the fit based on 
the statistical hadronization model with 
$T$ and $\gamma_S$. The third and fourth columns show the differences between 
data and model in units of standard error and in percentages, respectively.}\label{ee29}
\end{table}

\begin{table}[!h]\begin{center}
\begin{tabular}{|c|c|c|c|c|c|}
\hline
 &               & Experiment (E)       & Model (M)  & Residual & (M - E)/E  (\%)\\ 
\hline
$\pi^0$            &\cite{Aihara:1984mi} &      5.30 \tpm      0.70  &        6.37  &         1.5     &       20.2    \\ 
$\pi^+$            &\cite{Aihara:1986mv} &      5.35 \tpm      0.25  &        5.31  &       -0.15     &     -0.715    \\ 
$K^+$              &\cite{Aihara:1986mv} &     0.700 \tpm     0.050  &       0.760  &         1.2     &       8.55    \\ 
$K_S^0$            &\cite{Schellman:1984yz,Aihara:1984mk,Abachi:1989pr} &     0.691 \tpm     0.029  &       0.725  &         1.2     &       4.96    \\ 
$\eta$             &\cite{Abachi:1987qd,Wormser:1988ru} &     0.584 \tpm     0.075  &       0.643  &        0.78     &       10.0    \\ 
$\rho^0$           &\cite{Abachi:1989em} &     0.900 \tpm     0.050  &       0.727  &        -3.5     &      -19.2    \\ 
$K^{*+}$           &\cite{Abachi:1987wc} &     0.310 \tpm     0.030  &       0.231  &        -2.6     &      -25.6    \\ 
$K^{*0}$           &\cite{Abachi:1989em,Aihara:1984mk} &     0.281 \tpm     0.022  &       0.227  &        -2.4     &      -19.3    \\ 
$p$                &\cite{Aihara:1986mv} &     0.300 \tpm     0.050  &       0.292  &       -0.17     &      -2.82    \\ 
$\eta'$            &\cite{Wormser:1988ru} &      0.26 \tpm      0.10  &        0.07  &        -1.9     &      -75.0    \\ 
$\phi$             &\cite{Aihara:1984pw} &     0.084 \tpm     0.022  &       0.084  &      -0.022     &     -0.593    \\ 
$\Lambda$          &\cite{Aihara:1984wx,delaVaissiere:1984xg,Geld:1992si}&     0.0983 \tpm    0.0060  &      0.1024  &        0.69     &       4.24    \\ 
$\Xi^-$            &\cite{Klein:1986ws,Abachi:1987ac} &    0.0083 \tpm    0.0020  &      0.0068  &       -0.72     &      -17.5    \\ 
$\Sigma^{*+}$      &\cite{Abachi:1987ac} &    0.0083 \tpm    0.0024  &      0.0111  &         1.2     &       34.8    \\ 
$K_2^{*+}$         &\cite{Abachi:1987wc} &     0.045 \tpm     0.022  &       0.014  &        -1.4     &      -69.9    \\ 
$\Omega$           &\cite{Klein:1987fu} &    0.0070 \tpm    0.0036  &      0.0004  &        -1.8     &      -94.6    \\ 
\hline
\end{tabular}\end{center}
\caption{Hadron multiplicities in \epe~collisions at 29 GeV,compared to the outcomes of the fit based on Hawking-Unruh radiation model. The third and fourth columns show the differences between 
data and model in units of standard error and in percentages, respectively.}\label{ee29bh}
\end{table}

\begin{table}[!h]\begin{center}
\begin{tabular}{|c|c|c|c|c|c|}
\hline
&                & Experiment (E)       & Model (M)  & Residual & (M - E)/E  (\%)\\ 
\hline
$\pi^0$            &\cite{Braunschweig:1986hr,Pitzl:1989qy,Behrend:1989gn} &      6.31 \tpm      0.35  &        6.48  &        0.49     &       2.73    \\ 
$\pi^+$            &\cite{Braunschweig:1988hv} &      5.45 \tpm      0.25  &        5.42  &       -0.14     &     -0.621    \\ 
$K^+$              &\cite{Braunschweig:1988hv} &      0.88 \tpm      0.10  &        0.78  &       -0.98     &      -11.2    \\ 
$K_S^0$            &\cite{Bartel:1983qp,Braunschweig:1989wg,Behrend:1989ae} &     0.740 \tpm     0.017  &       0.746  &        0.33     &      0.759    \\ 
$\eta$             &\cite{Pitzl:1989qy,Behrend:1989gn} &     0.636 \tpm     0.080  &       0.661  &        0.32     &       4.06    \\ 
$\rho^0$           &\cite{Althoff:1984iz,Bartel:1984rh} &     0.756 \tpm     0.077  &       0.739  &       -0.23     &      -2.30    \\ 
$K^{*+}$           &\cite{Bartel:1984rh,Behrend:1989ae,Braunschweig:1989wg} &     0.361 \tpm     0.046  &       0.248  &        -2.4     &      -31.2    \\ 
$p$                &\cite{Bartel:1981sw,Braunschweig:1988hv} &     0.302 \tpm     0.033  &       0.300  &      -0.078     &     -0.838    \\ 
$\Lambda$          &\cite{Braunschweig:1988wh,Behrend:1989ae} &     0.108 \tpm     0.010  &       0.108  &      -0.042     &     -0.391    \\ 
$\Xi^-$            &\cite{Braunschweig:1988wh} &    0.0060 \tpm    0.0021  &      0.0079  &        0.90     &       31.5    \\ 
\hline
\end{tabular}\end{center}
\caption{Hadron multiplicities in \epe~collisions at 35 GeV, compared to the outcomes of the fit based on 
the statistical hadronization model with 
$T$ and $\gamma_S$. The third and fourth columns show the differences between 
data and model in units of standard error and in percentages, respectively.}\label{ee35}
\end{table}

\begin{table}[!h]\begin{center}
\begin{tabular}{|c|c|c|c|c|c|}
\hline
&                & Experiment (E)       & Model (M)  & Residual & (M - E)/E  (\%)\\ 
\hline
$\pi^0$            &\cite{Braunschweig:1986hr,Pitzl:1989qy,Behrend:1989gn} &      6.31 \tpm      0.35  &        6.46  &        0.41   &       2.31    \\ 
$\pi^+$            &\cite{Braunschweig:1988hv} &      5.45 \tpm      0.25  &        5.39  &       -0.25   &      -1.13    \\ 
$K^+$              &\cite{Braunschweig:1988hv} &      0.88 \tpm      0.10  &        0.78  &       -0.97   &      -11.0    \\ 
$K_S^0$            &\cite{Bartel:1983qp,Braunschweig:1989wg,Behrend:1989ae} &     0.740 \tpm     0.017  &       0.748  &        0.47   &       1.08    \\ 
$\eta$             &\cite{Pitzl:1989qy,Behrend:1989gn} &     0.636 \tpm     0.080  &       0.657  &        0.26   &       3.33    \\ 
$\rho^0$           &\cite{Althoff:1984iz,Bartel:1984rh} &     0.756 \tpm     0.077  &       0.735  &       -0.28   &      -2.82    \\ 
$K^{*+}$           &\cite{Bartel:1984rh,Behrend:1989ae,Braunschweig:1989wg} &     0.361 \tpm     0.046  &       0.239  &        -2.6   &      -33.6    \\ 
$p$                &\cite{Bartel:1981sw,Braunschweig:1988hv} &     0.302 \tpm     0.033  &       0.301  &      -0.042   &     -0.450    \\ 
$\Lambda$          &\cite{Braunschweig:1988wh,Behrend:1989ae} &     0.108 \tpm     0.010  &       0.108  &       0.028   &      0.261    \\ 
$\Xi^-$            &\cite{Braunschweig:1988wh} &    0.0060 \tpm    0.0021  &      0.0075  &        0.70   &       24.4    \\ 
\hline
\end{tabular}\end{center}
\caption{Hadron multiplicities in \epe~collisions at 35 GeV,compared to the outcomes of the fit based on Hawking-Unruh radiation model. 
The third and fourth columns show the differences between 
data and model in units of standard error and in percentages, respectively.}\label{ee35bh}
\end{table}

\begin{table}[!h]\begin{center}
\begin{tabular}{|c|c|c|c|c|c|}
\hline
&                & Experiment (E)       & Model (M)  & Residual & (M - E)/E  (\%)\\ 
\hline
$\pi^0$            &\cite{Braunschweig:1988hv,Pitzl:1989qy} &      6.66 \tpm      0.65  &        6.63  &      -0.055     &     -0.541    \\ 
$\pi^+$            &\cite{Braunschweig:1988hv} &      5.55 \tpm      0.25  &        5.55  &     -0.0021     &   -0.00955    \\ 
$K^+$              &\cite{Braunschweig:1989bp} &      0.96 \tpm      0.15  &        0.81  &        -1.0     &      -15.9    \\ 
$K_S^0$            &\cite{Braunschweig:1989wg} &     0.760 \tpm     0.035  &       0.775  &        0.43     &       2.01    \\ 
$K^{*+}$           &\cite{Braunschweig:1989wg} &     0.385 \tpm     0.094  &       0.264  &        -1.3     &      -31.3    \\ 
$\Lambda$          &\cite{Braunschweig:1988wh} &     0.128 \tpm     0.024  &       0.131  &        0.12     &       2.23    \\ 
\hline
\end{tabular}\end{center}
\caption{Hadron multiplicities in \epe~collisions at 43 GeV, compared to the outcomes of the fit based on 
the statistical hadronization model with 
$T$ and $\gamma_S$. The third and fourth columns show the differences between 
data and model in units of standard error and in percentages, respectively.}\label{ee43}
\end{table}

\begin{table}[!h]\begin{center}
\begin{tabular}{|c|c|c|c|c|c|}
\hline
&                & Experiment (E)       & Model (M)  & Residual & (M - E)/E  (\%)\\ 
\hline
$\pi^0$            &\cite{Braunschweig:1988hv,Pitzl:1989qy} &      6.66 \tpm      0.65  &        6.62  &      -0.065   &     -0.639    \\ 
$\pi^+$            &\cite{Braunschweig:1988hv} &      5.55 \tpm      0.25  &        5.54  &      -0.042   &     -0.188    \\ 
$K^+$              &\cite{Braunschweig:1989bp} &      0.96 \tpm      0.15  &        0.81  &        -1.0   &      -15.8    \\ 
$K_S^0$            &\cite{Braunschweig:1989wg} &     0.760 \tpm     0.035  &       0.777  &        0.47   &       2.18    \\ 
$K^{*+}$           &\cite{Braunschweig:1989wg} &     0.385 \tpm     0.094  &       0.255  &        -1.4   &      -33.6    \\ 
$\Lambda$          &\cite{Braunschweig:1988wh} &     0.128 \tpm     0.024  &       0.131  &        0.13   &       2.52    \\ 
\hline
\end{tabular}\end{center}
\caption{Hadron multiplicities in \epe~collisions at 43 GeV,compared to the outcomes of the fit based on 
Hawking-Unruh radiation model. The third and fourth columns show the differences between 
data and model in units of standard error and in percentages, respectively.}\label{ee43bh}
\end{table}

\begin{table}[!h]\begin{center}
\begin{tabular}{|c|c|c|c|c|c|}
\hline
&                & Experiment (E)       & Model (M)  & Residual & (M - E)/E  (\%)\\ 
\hline
$\pi^0$            &\cite{Acciarri:1994gza,Adam:1995rf,Barate:1996uh,Abbiendi:2000cv} &      9.61 \tpm      0.29  &       10.20  &         2.0   &       6.18    \\ 
$\pi^+$            &\cite{Akers:1994ez,Barate:1997ty,Abreu:1998vq,Abe:1998zs} &      8.50 \tpm      0.10  &        8.76  &         2.6   &       3.13    \\ 
$K^+$              &\cite{Akers:1994ez,Barate:1997ty,Abreu:1998vq,Abe:1998zs} &     1.127 \tpm     0.026  &       1.091  &        -1.4   &      -3.20    \\ 
$K_S^0$            &\cite{Abreu:1994rg,Acciarri:1994gza,Abe:1998zs,Abbiendi:2000cv,Barate:1999gb} &    1.0376 \tpm    0.0096  &      1.0507  &         1.4   &       1.26    \\ 
$\eta$             &\cite{Acciarri:1994gza,Abbiendi:2000cv,Heister:2001kp} &     1.059 \tpm     0.086  &       1.041  &       -0.21   &      -1.70    \\ 
$\rho^0$           &\cite{Abreu:1998nn,Krebs:1999iy} &      1.40 \tpm      0.13  &        1.19  &        -1.6   &      -15.0    \\ 
$\rho^+$           &\cite{Ackerstaff:1998ap} &      1.20 \tpm      0.22  &        1.14  &       -0.26   &      -4.66    \\ 
$\omega$           &\cite{Acciarri:1996tc,Ackerstaff:1998ap,Heister:2001kp} &     1.024 \tpm     0.059  &       1.014  &       -0.17   &     -0.997    \\ 
$K^{*+}$           &\cite{Acton:1992us,Abreu:1994rg,Barate:1996fi} &     0.357 \tpm     0.022  &       0.353  &       -0.16   &      -1.02    \\ 
$K^{*0}$           &\cite{Akers:1995wx,Abreu:1996sn,Barate:1996fi,Abe:1998zs} &     0.370 \tpm     0.013  &       0.346  &        -1.9   &      -6.33    \\ 
$p$                &\cite{Akers:1994ez,Abreu:1998vq,Abe:1998zs,Barate:1997ty} &     0.519 \tpm     0.018  &       0.564  &         2.5   &       8.76    \\ 
$\eta'$            &\cite{Acciarri:1996tc,Ackerstaff:1998ap} &     0.166 \tpm     0.047  &       0.106  &        -1.3   &      -36.1    \\ 
$f_0$              &\cite{Ackerstaff:1998ue,Abreu:1998nn,Krebs:1999iy} &    0.1555 \tpm    0.0085  &      0.0779  &        -9.1   &      -49.9    \\ 
$a_0^+$            &\cite{Ackerstaff:1998ap} &     0.135 \tpm     0.054  &       0.084  &       -0.95   &      -37.8    \\ 
$\phi$             &\cite{Abreu:1996sn,Barate:1996fi,Ackerstaff:1998ue,Abe:1998zs} &    0.0977 \tpm    0.0058  &      0.1150  &         3.0   &       17.7    \\ 
$\Lambda$          &\cite{Abreu:1993mm,Acciarri:1994gza,Alexander:1996qj,Abe:1998zs,Barate:1999gb} &    0.1943 \tpm    0.0038  &      0.1779  &        -4.3   &      -8.42    \\ 
$\Sigma^{+}$       &\cite{Alexander:1996qi,Acciarri:2000zf,Abreu:1995qx} &    0.0535 \tpm    0.0052  &      0.0415  &        -2.3   &      -22.4    \\ 
$\Sigma^{0}$       &\cite{Alexander:1996qi,Acciarri:2000zf,Adam:1996hw,Barate:1996fi} &    0.0389 \tpm    0.0041  &      0.0421  &        0.77   &       8.11    \\ 
$\Sigma^{-}$       &\cite{Alexander:1996qi,Abreu:2000nu} &    0.0410 \tpm    0.0037  &      0.0378  &       -0.85   &      -7.65    \\ 
$\Delta^{++}$      &\cite{Abreu:1995we,Alexander:1995gq} &     0.044 \tpm     0.017  &       0.090  &         2.7   &       105.    \\ 
$f_2$              &\cite{Ackerstaff:1998ue,Abreu:1998nn,Krebs:1999iy} &     0.188 \tpm     0.020  &       0.122  &        -3.4   &      -35.1    \\ 
$f_1$              &\cite{Abdallah:2003gu} &     0.165 \tpm     0.051  &       0.064  &        -2.0   &      -61.5    \\ 
$\Xi^-$            &\cite{Abreu:1995qx,Alexander:1996qj,Barate:1996fi} &   0.01319 \tpm   0.00050  &     0.01187  &        -2.6   &      -10.0    \\ 
$\Sigma^{*+}$      &\cite{Abreu:1995qx,Alexander:1996qj,Barate:1996fi} &    0.0118 \tpm    0.0011  &      0.0201  &         7.5   &       70.3    \\ 
$f'_1$             &\cite{Abdallah:2003gu} &     0.056 \tpm     0.012  &       0.010  &        -3.9   &      -82.6    \\ 
$K^{*}_{20}$       &\cite{Abreu:1998nn} &     0.036 \tpm     0.012  &       0.026  &       -0.91   &      -29.2    \\ 
$\Lambda(1520)$    &\cite{Alexander:1996qj,Abreu:2000nu} &    0.0112 \tpm    0.0014  &      0.0109  &       -0.22   &      -2.73    \\ 
$f'_2$             &\cite{Abreu:1998nn} &    0.0120 \tpm    0.0058  &      0.0100  &       -0.34   &      -16.6    \\ 
$\Xi^{*0}$         &\cite{Abreu:1995qx,Alexander:1996qj,Barate:1996fi} &   0.00289 \tpm   0.00050  &     0.00417  &         2.6   &       44.4    \\ 
$\Omega$           &\cite{Adam:1996hw,Alexander:1996qj,Barate:1996fi} &   0.00062 \tpm   0.00010  &     0.00081  &         1.9   &       31.0    \\ 
\hline
\end{tabular}\end{center}
\caption{Hadron multiplicities in \epe~collisions at 91 GeV, compared to the outcomes of the fit based on 
the statistical hadronization model with 
$T$ and $\gamma_S$. The third and fourth columns show the differences between 
data and model in units of standard error and in percentages, respectively.}\label{ee91}
\end{table}

\newpage
\begin{table}[!h]\begin{center}
\begin{tabular}{|c|c|c|c|c|c|}
\hline
&                & Experiment (E)       & Model (M)  & Residual & (M - E)/E  (\%)\\ 
\hline
$\pi^0$            &\cite{Acciarri:1994gza,Adam:1995rf,Barate:1996uh,Abbiendi:2000cv} &      9.61 \tpm      0.29  &       10.01  &         1.4   &       4.18    \\ 
$\pi^+$            &\cite{Akers:1994ez,Barate:1997ty,Abreu:1998vq,Abe:1998zs} &      8.50 \tpm      0.10  &        8.57  &        0.77   &      0.918    \\ 
$K^+$              &\cite{Akers:1994ez,Barate:1997ty,Abreu:1998vq,Abe:1998zs} &     1.127 \tpm     0.026  &       1.131  &        0.16   &      0.373    \\ 
$K_S^0$            &\cite{Abreu:1994rg,Acciarri:1994gza,Abe:1998zs,Abbiendi:2000cv,Barate:1999gb} &    1.0376 \tpm    0.0096  &      1.0901  &         5.5   &       5.05    \\ 
$\eta$             &\cite{Acciarri:1994gza,Abbiendi:2000cv,Heister:2001kp} &     1.059 \tpm     0.086  &       1.026  &       -0.38   &      -3.12    \\ 
$\rho^0$           &\cite{Abreu:1998nn,Krebs:1999iy} &      1.40 \tpm      0.13  &        1.15  &        -1.9   &      -17.4    \\ 
$\rho^+$           &\cite{Ackerstaff:1998ap} &      1.20 \tpm      0.22  &        1.11  &       -0.42   &      -7.53    \\ 
$\omega$           &\cite{Acciarri:1996tc,Ackerstaff:1998ap,Heister:2001kp} &     1.024 \tpm     0.059  &       0.982  &       -0.71   &      -4.08    \\ 
$K^{*+}$           &\cite{Acton:1992us,Abreu:1994rg,Barate:1996fi} &     0.357 \tpm     0.022  &       0.345  &       -0.54   &      -3.36    \\ 
$K^{*0}$           &\cite{Akers:1995wx,Abreu:1996sn,Barate:1996fi,Abe:1998zs} &     0.370 \tpm     0.013  &       0.338  &        -2.5   &      -8.62    \\ 
$p$                &\cite{Akers:1994ez,Abreu:1998vq,Abe:1998zs,Barate:1997ty} &     0.519 \tpm     0.018  &       0.548  &         1.6   &       5.67    \\ 
$\eta'$            &\cite{Acciarri:1996tc,Ackerstaff:1998ap} &     0.166 \tpm     0.047  &       0.093  &        -1.6   &      -43.8    \\ 
$f_0$              &\cite{Ackerstaff:1998ue,Abreu:1998nn,Krebs:1999iy} &    0.1555 \tpm    0.0085  &      0.0751  &        -9.5   &      -51.7    \\ 
$a_0^+$            &\cite{Ackerstaff:1998ap} &     0.135 \tpm     0.054  &       0.081  &        -1.0   &      -40.0    \\ 
$\phi$             &\cite{Abreu:1996sn,Barate:1996fi,Ackerstaff:1998ue,Abe:1998zs} &    0.0977 \tpm    0.0058  &      0.1048  &         1.2   &       7.19    \\ 
$\Lambda$          &\cite{Abreu:1993mm,Acciarri:1994gza,Alexander:1996qj,Abe:1998zs,Barate:1999gb} &    0.1943 \tpm    0.0038  &      0.1826  &        -3.1   &      -6.04    \\ 
$\Sigma^{+}$       &\cite{Alexander:1996qi,Acciarri:2000zf,Abreu:1995qx} &    0.0535 \tpm    0.0052  &      0.0424  &        -2.1   &      -20.7    \\ 
$\Sigma^{0}$       &\cite{Alexander:1996qi,Acciarri:2000zf,Adam:1996hw,Barate:1996fi} &    0.0389 \tpm    0.0041  &      0.0430  &        1.00   &       10.5    \\ 
$\Sigma^{-}$       &\cite{Alexander:1996qi,Abreu:2000nu} &    0.0410 \tpm    0.0037  &      0.0388  &       -0.59   &      -5.31    \\ 
$\Delta^{++}$      &\cite{Abreu:1995we,Alexander:1995gq} &     0.044 \tpm     0.017  &       0.086  &         2.5   &       95.0    \\ 
$f_2$              &\cite{Ackerstaff:1998ue,Abreu:1998nn,Krebs:1999iy} &     0.188 \tpm     0.020  &       0.115  &        -3.7   &      -38.9    \\ 
$f_1$              &\cite{Abdallah:2003gu} &     0.165 \tpm     0.051  &       0.061  &        -2.0   &      -63.2    \\ 
$\Xi^-$            &\cite{Abreu:1995qx,Alexander:1996qj,Barate:1996fi} &   0.01319 \tpm   0.00050  &     0.01204  &        -2.3   &      -8.72    \\ 
$\Sigma^{*+}$      &\cite{Abreu:1995qx,Alexander:1996qj,Barate:1996fi} &    0.0118 \tpm    0.0011  &      0.0204  &         7.8   &       72.8    \\ 
$f'_1$             &\cite{Abdallah:2003gu} &     0.056 \tpm     0.012  &       0.007  &        -4.1   &      -87.3    \\ 
$K^{*}_{20}$       &\cite{Abreu:1998nn} &     0.036 \tpm     0.012  &       0.021  &        -1.3   &      -41.7    \\ 
$\Lambda(1520)$    &\cite{Alexander:1996qj,Abreu:2000nu} &    0.0112 \tpm    0.0014  &      0.0106  &       -0.45   &      -5.55    \\ 
$f'_2$             &\cite{Abreu:1998nn} &    0.0120 \tpm    0.0058  &      0.0068  &       -0.89   &      -43.1    \\ 
$\Xi^{*0}$         &\cite{Abreu:1995qx,Alexander:1996qj,Barate:1996fi} &   0.00289 \tpm   0.00050  &     0.00423  &         2.7   &       46.4    \\ 
$\Omega$           &\cite{Adam:1996hw,Alexander:1996qj,Barate:1996fi} &   0.00062 \tpm   0.00010  &     0.00071  &        0.85   &       13.8    \\ 
\hline
\end{tabular}\end{center}
\caption{Hadron multiplicities in \epe~collisions at 91 GeV,compared to the outcomes of the fit based on 
Hawking-Unruh radiation model. The third and fourth columns show the differences between 
data and model in units of standard error and in percentages, respectively.}\label{ee91bh}
\end{table}

\begin{table}[!h]\begin{center}
\begin{tabular}{|c|c|c|c|c|c|}
\hline
&                & Experiment (E)       & Model (M)  & Residual & (M - E)/E  (\%)\\ 
\hline
$\pi^+$            &\cite{Abreu:2000gw} &      9.92 \tpm      0.26  &        9.94  &       0.063   &      0.167    \\ 
$K^+$              &\cite{Abreu:2000gw} &      1.30 \tpm      0.15  &        1.29  &      -0.064   &     -0.714    \\ 
$K_S^0$            &\cite{Abreu:2000gw} &      1.25 \tpm      0.12  &        1.25  &      -0.070   &     -0.663    \\ 
$p$                &\cite{Abreu:2000gw} &      0.78 \tpm      0.13  &        0.75  &       -0.19   &      -3.26    \\ 
$\Lambda$          &\cite{Abreu:2000gw} &     0.250 \tpm     0.038  &       0.256  &        0.15   &       2.27    \\ 
\hline
\end{tabular}\end{center}
\caption{Hadron multiplicities in \epe~collisions at 133 GeV, compared to the outcomes of the fit based on 
the statistical hadronization model with 
$T$ and $\gamma_S$. The third and fourth columns show the differences between 
data and model in units of standard error and in percentages, respectively.}\label{ee133}
\end{table}

\begin{table}[!h]\begin{center}
\begin{tabular}{|c|c|c|c|c|c|}
\hline
&                & Experiment (E)       & Model (M)  & Residual & (M - E)/E  (\%)\\ 
\hline
$\pi^+$            &\cite{Abreu:2000gw} &      9.92 \tpm      0.26  &        9.94  &       0.059   &      0.157    \\ 
$K^+$              &\cite{Abreu:2000gw} &      1.30 \tpm      0.15  &        1.29  &      -0.060   &     -0.669    \\ 
$K_S^0$            &\cite{Abreu:2000gw} &      1.25 \tpm      0.12  &        1.25  &      -0.056   &     -0.533    \\ 
$p$                &\cite{Abreu:2000gw} &      0.78 \tpm      0.13  &        0.76  &       -0.17   &      -2.92    \\ 
$\Lambda$          &\cite{Abreu:2000gw} &     0.250 \tpm     0.038  &       0.255  &        0.13   &       1.92    \\ 
\hline
\end{tabular}\end{center}
\caption{Hadron multiplicities in \epe~collisions at 133 GeV,compared to the outcomes of the fit based on 
Hawking-Unruh radiation model. The third and fourth columns show the differences between 
data and model in units of standard error and in percentages, respectively.}\label{ee133bh}
\end{table}

\begin{table}[!h]\begin{center}
\begin{tabular}{|c|c|c|c|c|c|}
\hline
&                & Experiment (E)       & Model (M)  & Residual & (M - E)/E  (\%)\\ 
\hline
$\pi^+$            &\cite{Abreu:2000gw} &     10.38 \tpm      0.38  &       10.37  &    -0.00096   &   -0.00355    \\ 
$K^+$              &\cite{Abreu:2000gw} &      1.44 \tpm      0.30  &        1.39  &       -0.15   &      -3.12    \\ 
$K_S^0$            &\cite{Abreu:2000gw} &      1.32 \tpm      0.18  &        1.34  &       0.094   &       1.31    \\ 
$p$                &\cite{Abreu:2000gw} &      0.60 \tpm      0.24  &        0.60  &     -0.0010   &    -0.0414    \\ 
\hline
\end{tabular}\end{center}
\caption{Hadron multiplicities in \epe~collisions at 161 GeV, compared to the outcomes of the fit based on 
the statistical hadronization model with 
$T$ and $\gamma_S$. The third and fourth columns show the differences between 
data and model in units of standard error and in percentages, respectively.}\label{ee161}
\end{table}

\begin{table}[!h]\begin{center}
\begin{tabular}{|c|c|c|c|c|c|}
\hline
  &              & Experiment (E)       & Model (M)  & Residual & (M - E)/E  (\%)\\ 
\hline
$\pi^+$            &\cite{Abreu:2000gw} &     10.38 \tpm      0.38  &       10.37  &     -0.0012   &   -0.00431    \\ 
$K^+$              &\cite{Abreu:2000gw} &      1.44 \tpm      0.30  &        1.39  &       -0.15   &      -3.18    \\ 
$K_S^0$            &\cite{Abreu:2000gw} &      1.32 \tpm      0.18  &        1.34  &       0.097   &       1.34    \\ 
$p$                &\cite{Abreu:2000gw} &      0.60 \tpm      0.24  &        0.60  &     -0.0011   &    -0.0436    \\ 
\hline
\end{tabular}\end{center}
\caption{Hadron multiplicities in \epe~collisions at 161 GeV,compared to the outcomes of the fit based on 
Hawking-Unruh radiation model. The third and fourth columns show the differences between 
data and model in units of standard error and in percentages, respectively.}\label{ee161bh}
\end{table}

\begin{table}[!h]\begin{center}
\begin{tabular}{|c|c|c|c|c|c|}
\hline
&                & Experiment (E)       & Model (M)  & Residual & (M - E)/E  (\%)\\ 
\hline
$\pi^+$            &\cite{Abreu:2000gw} &     10.89 \tpm      0.29  &       10.87  &      -0.081   &     -0.216    \\ 
$K^+$              &\cite{Abreu:2000gw} &      1.42 \tpm      0.20  &        1.03  &        -2.0   &      -27.2    \\ 
$K_S^0$            &\cite{Abreu:2000gw} &     0.905 \tpm     0.086  &       0.995  &         1.0   &       9.97    \\ 
$p$                &\cite{Abreu:2000gw} &      0.66 \tpm      0.19  &        0.71  &        0.25   &       7.13    \\ 
$\Lambda$          &\cite{Abreu:2000gw} &     0.165 \tpm     0.025  &       0.161  &       -0.15   &      -2.35    \\ 
\hline
\end{tabular}\end{center}
\caption{Hadron multiplicities in \epe~collisions at 183 GeV, compared to the outcomes of the fit based on 
the statistical hadronization model with 
$T$ and $\gamma_S$. The third and fourth columns show the differences between 
data and model in units of standard error and in percentages, respectively.}\label{ee183}
\end{table}

\begin{table}[!h]\begin{center}
\begin{tabular}{|c|c|c|c|c|c|}
\hline
&                & Experiment (E)       & Model (M)  & Residual & (M - E)/E  (\%)\\ 
\hline
$\pi^+$            &\cite{Abreu:2000gw} &     10.89 \tpm      0.29  &       10.86  &       -0.12   &     -0.326    \\ 
$K^+$              &\cite{Abreu:2000gw} &      1.42 \tpm      0.20  &        1.03  &        -1.9   &      -27.0    \\ 
$K_S^0$            &\cite{Abreu:2000gw} &     0.905 \tpm     0.086  &       0.999  &         1.1   &       10.4    \\ 
$p$                &\cite{Abreu:2000gw} &      0.66 \tpm      0.19  &        0.73  &        0.36   &       10.5    \\ 
$\Lambda$          &\cite{Abreu:2000gw} &     0.165 \tpm     0.025  &       0.160  &       -0.20   &      -3.08    \\ 
\hline
\end{tabular}\end{center}
\caption{Hadron multiplicities in \epe~collisions at 183 GeV,compared to the outcomes of the fit based on 
Hawking-Unruh radiation model. The third and fourth columns show the differences between 
data and model in units of standard error and in percentages, respectively.}\label{ee183bh}
\end{table}

\clearpage

\begin{table}[!h]\begin{center}
\begin{tabular}{|c|c|c|c|c|c|}
\hline
&                & Experiment (E)       & Model (M)  & Residual & (M - E)/E  (\%)\\ 
\hline
$\pi^+$            &\cite{Abreu:2000gw} &     11.10 \tpm      0.26  &       11.06  &       -0.15   &     -0.344    \\ 
$K^+$              &\cite{Abreu:2000gw} &      1.57 \tpm      0.16  &        1.21  &        -2.3   &      -23.2    \\ 
$K_S^0$            &\cite{Abreu:2000gw} &     1.060 \tpm     0.078  &       1.169  &         1.4   &       10.3    \\ 
$p$                &\cite{Abreu:2000gw} &      0.59 \tpm      0.22  &        0.72  &        0.54   &       20.3    \\ 
$\Lambda$          &\cite{Abreu:2000gw} &     0.200 \tpm     0.021  &       0.196  &       -0.20   &      -2.16    \\ 
\hline
\end{tabular}\end{center}
\caption{Hadron multiplicities in \epe~collisions at 189 GeV, compared to the outcomes of the fit based on 
the statistical hadronization model with 
$T$ and $\gamma_S$. The third and fourth columns show the differences between 
data and model in units of standard error and in percentages, respectively.}\label{ee189}
\end{table}

\begin{table}[!h]\begin{center} 
\begin{tabular}{|c|c|c|c|c|c|}
\hline
&                & Experiment (E)       & Model (M)  & Residual & (M - E)/E  (\%)\\ 
\hline
$\pi^+$            &\cite{Abreu:2000gw} &     11.10 \tpm      0.26  &       11.05  &       -0.18   &     -0.422    \\ 
$K^+$              &\cite{Abreu:2000gw} &      1.57 \tpm      0.16  &        1.21  &        -2.3   &      -23.2    \\ 
$K_S^0$            &\cite{Abreu:2000gw} &     1.060 \tpm     0.078  &       1.171  &         1.4   &       10.5    \\ 
$p$                &\cite{Abreu:2000gw} &      0.59 \tpm      0.22  &        0.74  &        0.64   &       23.7    \\ 
$\Lambda$          &\cite{Abreu:2000gw} &     0.200 \tpm     0.021  &       0.195  &       -0.22   &      -2.31    \\ 
\hline
\end{tabular}\end{center}
\caption{Hadron multiplicities in \epe~collisions at 189 GeV,compared to the outcomes of the fit based on 
Hawking-Unruh radiation model. 
The third and fourth columns show the differences between 
data and model in units of standard error and in percentages, respectively.}\label{ee189bh}
\end{table}


\clearpage


\begin{thebibliography}{99}

\bibitem{Beca-e}
 F.\ Becattini, \ZP C69 (1996) 485.

\bibitem{erice}
 F.\ Becattini, {\em Universality of thermal hadron production in $pp$, 
  $p\bar p$ and $e^+e^-$ collisions}, in 
  {\em Universality features in multihadron production and the leading effect},
  ,Singapore, World Scientific, 1998 p. 74-104, arXiv:hep-ph/9701275.

\bibitem{Beca-h}
 F.\ Becattini and G.\ Passaleva, \EP 23 (2002) 551.

\bibitem{Beca-p}
 F.\ Becattini and U.\ Heinz, \ZP C76 (1997) 268.
 
\bibitem{Beca-hi} 
 J.\ Cleymans et al., \PL B 242 (1990) 111;\\
 J.\ Cleymans and H.\ Satz, \ZP C57 (1993) 135;\\
 K.\ Redlich et al., \NP A 566 (1994) 391;\\
 P. Braun-Munzinger et al., \PL B344 (1995) 43;\\
 F.\ Becattini, M.\ Gazdzicki and J.\ Sollfrank, \EP 5 (1998) 143;\\ 
 F.\ Becattini et al., \PR C64 (2001) 024901;\\
 P.\ Braun-Munzinger, K.\ Redlich and J.\ Stachel, in {\sl Quark-Gluon
 Plasma 3}, R.\ C.\ Hwa and X.-N\ Wang (Eds.), World Scientific,
 Singapore 2003.

\bibitem{Beca-Biele}
 F.\ Becattini, \NP A702 (2001) 336c.

\bibitem{Raf} J.\ Letessier, J.\ Rafelski and A.\ Tounsi, \PR C64 (1994) 406.

\bibitem{gamma} See e.g. F.~Becattini, J.~Manninen and M.~Gazdzicki,
  Phys.\ Rev.\  C {\bf 73} (2006) 044905; 
  P.~Braun-Munzinger, D.~Magestro, K.~Redlich and J.~Stachel, 
  Phys.\ Lett.\  B {\bf 518} (2001) 41.

\bibitem{diba}  
 U. Heinz, \NP A 661 (1999) 140;\\
 R. Stock, \PL B 456 (1999) 277;\\ 
 A. Bialas, \PL B 466  (1999) 301;\\
 H. Satz, Nucl. Phys. Proc. Suppl. {\bf 94} (2001) 204; \\
 J. Hormuzdiar,S. D. H. Hsu, G. Mahlon Int. J. Mod. Phys. E 12 (2003) 649;
 V. Koch, \NP A 715(2003) 108 ; \\
 L. McLerran, arXiv:hep-ph/0311028;\\
 Y. Dokshitzer, Acta Phys. Polon. B {\bf 36} (2005) 361; \\ 
 F. Becattini, J. Phys. Conf. Ser. {\bf 5} (2005) 175. 

\bibitem{CKS} P.\ Castorina, D.\ Kharzeev and H.\ Satz, \EP 52 (2007) 187.

\bibitem{Hawk} S.\ W.\ Hawking, Comm.\ Math.\ Phys.\ 43 (1975) 199. 

\bibitem{Un} W.\ G.\ Unruh, \PR D 14 (1976) 870.

\bibitem{Schwinger} J.\ Schwinger, \PR 82 (1951) 664.

\bibitem{KT} 
R.\ Brout, R.\ Parentani and Ph.\ Spindel, \NP B 353 (1991) 209;\\
P.\ Parentani and S.\ Massar, \PR D 55 (1997) 3603;\\
K.\ Srinivasan and T.\ Padmanabhan, \PR D60 (1999) 024007;\\
D.~Kharzeev and K.~Tuchin, Nucl.\ Phys.\  A {\bf 753}, 316 (2005);\\
Sang Pyo Kim, arXiv:0709.4313 [hep-th] 2007

\bibitem{meaning}
 F. Becattini, J. Phys. Conf. Ser. {\bf 5} (2005) 175.

\bibitem{Beca-micro}
 F.~Becattini and L.~Ferroni, Eur.\ Phys.\ J.\  C {\bf 35} (2004) 243;
 F.~Becattini and L.~Ferroni, Eur.\ Phys.\ J.\  C {\bf 38} (2004) 225.

\bibitem{chemfact}
  A.~Keranen and F.~Becattini, Phys.\ Rev.\  C {\bf 65} (2002) 044901,
  Erratum-ibid.\  C {\bf 68} (2003) 059901.

\bibitem{bj} J.\ D.\ Bjorken, Lecture Notes in Physics (Springer) 56 
(1976) 93.

\bibitem{nus} A.\ Casher, H.\ Neuberger and S.\ Nussinov, \PR D20 (1979) 179.

\bibitem{pdg} W.-M.\ Yao et al.\ (2006 Review of Particle Physics),
 J.\ Phys.\ G 33 (2006) 1.

\bibitem{pythia}
 T.\ Sjostrand et al.\ Comp.\ Phys.\ Comm.\ {\bf 135} (2001) 238.  

\bibitem{JOS} S. Jacobs, M.\ G.\ Olsson and C.\ Suchyta, \PR D 33 (1986) 3338.

\bibitem{Nora} F.\ J.\ Yndurain, {\sl Theory of Quark and Gluon 
Interactions}, Springer Verlag Berlin, 1999;\\
N.\ Brambilla et al., CERN Yellow Report CERN-2005-005.

\bibitem{BBC} C.\ Aubin et al. (MILC Collaboration), \PR D70 (2004) 094505;\\
A.\ Gray et al., \PR D72 (2005) 0894507,\\
M.\ Cheng et al., arXiv:hep-lat/0608013.

\bibitem{santorini}
F.~Becattini, J.\ Phys.\ G {\bf 23} (1997) 1933.

\bibitem{P-W} M.\ K.\ Parikh and F.\ Wilczek, \PRL 85 (2000) 5042. 

\bibitem{pbmee} A.~Andronic, F.~Beutler, P.~Braun-Munzinger, K.~Redlich and J.~Stachel,
  arXiv:0804.4132.






\bibitem{99bg}
  R.~Barate {\it et al.}  [ALEPH Collaboration],
  Eur.\ Phys.\ J.\  C {\bf 16} (2000) 597.

\bibitem{97ki}
  K.~Ackerstaff {\it et al.}  [OPAL Collaboration],
  Eur.\ Phys.\ J.\  C {\bf 1} (1998) 439.

\bibitem{99vx}
  P.~Abreu {\it et al.}  [DELPHI Collaboration],
  Eur.\ Phys.\ J.\  C {\bf 12} (2000) 209.

\bibitem{98as}
  K.~Ackerstaff {\it et al.}  [OPAL Collaboration],
  Eur.\ Phys.\ J.\  C {\bf 5} (1998) 1.

\bibitem{alconf}
 TheALEPH Collaboration, 
 ``Production of D1 and D*2 mesons in hadronic Z decays",
 preprint ALEPH 98-047 CONF 98-021 (1998). 

\bibitem{97vc}
  K.~Ackerstaff {\it et al.}  [OPAL Collaboration],
  Z.\ Phys.\  C {\bf 76} (1997) 425.

\bibitem{01nj}
  A.~Heister {\it et al.}  [ALEPH Collaboration],
  Phys.\ Lett.\  B {\bf 526} (2002) 34.


\bibitem{Abbaneo:2001bv}
  The ALEPH, DELPHI, L3, OPAL and CDF Collaborations, 
 ``Combined results on b-hadron production rates and decay properties,''
  arXiv:hep-ex/0112028.

\bibitem{96gz}
  K.~Ackerstaff {\it et al.}  [OPAL Collaboration],
  Z.\ Phys.\  C {\bf 74} (1997) 413.

\bibitem{95mt}
  D.~Buskulic {\it et al.}  [ALEPH Collaboration],
  Z.\ Phys.\  C {\bf 69} (1996) 393.

\bibitem{95ky}
  P.~Abreu {\it et al.}  [DELPHI Collaboration],
  Z.\ Phys.\  C {\bf 68} (1995) 353.

\bibitem{94qv}
  M.~Acciarri {\it et al.}  [L3 Collaboration],
  Phys.\ Lett.\  B {\bf 345} (1995) 589.

\bibitem{98cq}
  R.~Barate {\it et al.}  [ALEPH Collaboration],
  Phys.\ Lett.\  B {\bf 425} (1998) 215.

\bibitem{delconf}
  Z. Albrecht {\it et al.} [DELPHI Collaboration]
  ``A study of excited b-hadron states with the DELPHI detector at LEP",
  preprint DELPHI-2004-025 CONF 700 (2004), presented at ICHEP 2004.

\bibitem{94fz}
  R.~Akers {\it et al.}  [OPAL Collaboration],
  Z.\ Phys.\  C {\bf 66} (1995) 19.

\bibitem{Bartel:1985wn}
  W.~Bartel {\it et al.}  [JADE Collaboration],
  Z.\ Phys.\  C {\bf 28} (1985) 343.

\bibitem{Althoff:1984iz}
  M.~Althoff {\it et al.}  [TASSO Collaboration],
  Z.\ Phys.\  C {\bf 27} (1985) 27.

\bibitem{Bartel:1983qp}
  W.~Bartel {\it et al.}  [JADE Collaboration],
  Z.\ Phys.\  C {\bf 20} (1983) 187.

\bibitem{Braunschweig:1989wg}
  W.~Braunschweig {\it et al.}  [TASSO Collaboration],
  Z.\ Phys.\  C {\bf 47} (1990) 167.

\bibitem{Aihara:1984mi}
  H.~Aihara {\it et al.}  [TPC/Two Gamma Collaboration],
  Z.\ Phys.\  C {\bf 27} (1985) 187.

\bibitem{Aihara:1986mv}
  H.~Aihara {\it et al.}  [TPC/Two Gamma Collaboration],
  Phys.\ Lett.\  B {\bf 184} (1987) 299.

\bibitem{Schellman:1984yz}
  H.~Schellman {\it et al.},
  Phys.\ Rev.\  D {\bf 31} (1985) 3013.

\bibitem{Aihara:1984mk}
  H.~Aihara {\it et al.}  [TPC/Two Gamma Collaboration],
  Phys.\ Rev.\ Lett.\  {\bf 53} (1984) 2378.

\bibitem{Abachi:1989pr}
  S.~Abachi {\it et al.}  [HRS Collaboration],
  Phys.\ Rev.\  D {\bf 41} (1990) 2045.

\bibitem{Abachi:1987qd}
  S.~Abachi {\it et al.}  [HRS Collaboration],
  Phys.\ Lett.\  B {\bf 205} (1988) 111.

\bibitem{Wormser:1988ru}
  G.~Wormser {\it et al.},
  Phys.\ Rev.\ Lett.\  {\bf 61} (1988) 1057.

\bibitem{Abachi:1989em}
  S.~Abachi {\it et al.},
  Phys.\ Rev.\  D {\bf 40} (1989) 706.

\bibitem{Abachi:1987wc}
  S.~Abachi {\it et al.},
  Phys.\ Lett.\  B {\bf 199} (1987) 151.

\bibitem{Aihara:1984pw}
  H.~Aihara {\it et al.}  [TPC/Two Gamma Collaboration],
  Phys.\ Rev.\ Lett.\  {\bf 52} (1984) 2201.

\bibitem{Aihara:1984wx}
  H.~Aihara {\it et al.}  [TPC/Two Gamma Collaboration],
  Phys.\ Rev.\ Lett.\  {\bf 54} (1985) 274.

\bibitem{delaVaissiere:1984xg}
  C.~de la Vaissiere {\it et al.},
  Phys.\ Rev.\ Lett.\  {\bf 54} (1985) 2071
  [Erratum-ibid.\  {\bf 55} (1985) 263].

\bibitem{Geld:1992si}
  T.~L.~Geld {\it et al.}  [HRS Collaboration],
  Phys.\ Rev.\  D {\bf 45} (1992) 3949.

\bibitem{Klein:1986ws}
  S.~Klein {\it et al.},
  Phys.\ Rev.\ Lett.\  {\bf 58} (1987) 644.

\bibitem{Abachi:1987ac}
  S.~Abachi {\it et al.},
  Phys.\ Rev.\ Lett.\  {\bf 58} (1987) 2627
  [Erratum-ibid.\  {\bf 59} (1987) 2388].

\bibitem{Klein:1987fu}
  S.~Klein {\it et al.},
  Phys.\ Rev.\ Lett.\  {\bf 59} (1987) 2412.

\bibitem{Braunschweig:1986hr}
  W.~Braunschweig {\it et al.}  [TASSO Collaboration],
  Z.\ Phys.\  C {\bf 33} (1986) 13.

\bibitem{Pitzl:1989qy}
  D.~Pitzl {\it et al.}  [JADE Collaboration],
  Z.\ Phys.\  C {\bf 46} (1990) 1
  [Erratum-ibid.\  C {\bf 47} (1990) 676].

\bibitem{Behrend:1989gn}
  H.~J.~Behrend {\it et al.}  [CELLO Collaboration],
  Z.\ Phys.\  C {\bf 47} (1990) 1.

\bibitem{Braunschweig:1988hv}
  W.~Braunschweig {\it et al.}  [TASSO Collaboration],
  Z.\ Phys.\  C {\bf 42} (1989) 189.

\bibitem{Behrend:1989ae}
  H.~J.~Behrend {\it et al.}  [CELLO Collaboration],
  Z.\ Phys.\  C {\bf 46} (1990) 397.

\bibitem{Bartel:1984rh}
  W.~Bartel {\it et al.}  [JADE Collaboration],
  Phys.\ Lett.\  B {\bf 145} (1984) 441.

\bibitem{Bartel:1981sw}
  W.~Bartel {\it et al.}  [JADE Collaboration],
  Phys.\ Lett.\  B {\bf 104} (1981) 325.

\bibitem{Braunschweig:1988wh}
  W.~Braunschweig {\it et al.}  [TASSO Collaboration],
  Z.\ Phys.\  C {\bf 45} (1989) 209.

\bibitem{Braunschweig:1989bp}
  W.~Braunschweig {\it et al.}  [TASSO Collaboration],
  Z.\ Phys.\  C {\bf 45} (1989) 193.

\bibitem{Acciarri:1994gza}
  M.~Acciarri {\it et al.}  [L3 Collaboration],
  Phys.\ Lett.\  B {\bf 328} (1994) 223.

\bibitem{Adam:1995rf}
  W.~Adam {\it et al.}  [DELPHI Collaboration],
  Z.\ Phys.\  C {\bf 69} (1996) 561.

\bibitem{Barate:1996uh}
  R.~Barate {\it et al.}  [ALEPH Collaboration],
  Z.\ Phys.\  C {\bf 74} (1997) 451.

\bibitem{Abbiendi:2000cv}
  G.~Abbiendi {\it et al.}  [OPAL Collaboration],
  Eur.\ Phys.\ J.\  C {\bf 17} (2000) 373.

\bibitem{Akers:1994ez}
  R.~Akers {\it et al.}  [OPAL Collaboration],
  Z.\ Phys.\  C {\bf 63} (1994) 181.

\bibitem{Barate:1997ty}
  R.~Barate {\it et al.}  [ALEPH Collaboration],
  Eur.\ Phys.\ J.\  C {\bf 5} (1998) 205.

\bibitem{Abreu:1998vq}
  P.~Abreu {\it et al.}  [DELPHI Collaboration],
  Eur.\ Phys.\ J.\  C {\bf 5} (1998) 585.

\bibitem{Abe:1998zs}
  K.~Abe {\it et al.}  [SLD Collaboration],
  Phys.\ Rev.\  D {\bf 59} (1999) 052001.

\bibitem{Abreu:1994rg}
  P.~Abreu {\it et al.}  [DELPHI Collaboration],
  Z.\ Phys.\  C {\bf 65} (1995) 587.

\bibitem{Akers:1995ay}
  R.~Akers {\it et al.}  [OPAL Collaboration],
  Z.\ Phys.\  C {\bf 67} (1995) 389.

\bibitem{Barate:1996fi}
  R.~Barate {\it et al.}  [ALEPH Collaboration],
  Phys.\ Rept.\  {\bf 294} (1998) 1.

\bibitem{Barate:1999gb}
  R.~Barate {\it et al.}  [ALEPH Collaboration],
  Eur.\ Phys.\ J.\  C {\bf 16} (2000) 613.

\bibitem{Heister:2001kp}
  A.~Heister {\it et al.}  [ALEPH Collaboration],
  Phys.\ Lett.\  B {\bf 528} (2002) 19.

\bibitem{Abreu:1998nn}
  P.~Abreu {\it et al.}  [DELPHI Collaboration],
  Phys.\ Lett.\  B {\bf 449} (1999) 364.

\bibitem{Krebs:1999iy}
  W.~Krebs  [ALEPH Collaboration],
  ALEPH-99-057.

\bibitem{Ackerstaff:1998ap}
  K.~Ackerstaff {\it et al.}  [OPAL Collaboration],
  Eur.\ Phys.\ J.\  C {\bf 5} (1998) 411.

\bibitem{Acciarri:1996tc}
  M.~Acciarri {\it et al.}  [L3 Collaboration],
  Phys.\ Lett.\  B {\bf 393} (1997) 465.

\bibitem{Acton:1992us}
  P.~D.~Acton {\it et al.}  [OPAL Collaboration],
  Phys.\ Lett.\  B {\bf 305} (1993) 407.

\bibitem{Akers:1995wx}
  R.~Akers {\it et al.}  [OPAL Collaboration],
  Z.\ Phys.\  C {\bf 68} (1995) 1.

\bibitem{Abreu:1996sn}
  P.~Abreu {\it et al.}  [DELPHI Collaboration],
  Z.\ Phys.\  C {\bf 73} (1996) 61.

\bibitem{Ackerstaff:1998ue}
  K.~Ackerstaff {\it et al.}  [OPAL Collaboration],
  Eur.\ Phys.\ J.\  C {\bf 4} (1998) 19.

\bibitem{Abreu:1993mm}
  P.~Abreu {\it et al.}  [DELPHI Collaboration],
  Phys.\ Lett.\  B {\bf 318} (1993) 249.

\bibitem{Alexander:1996qj}
  G.~Alexander {\it et al.}  [OPAL Collaboration],
  Z.\ Phys.\  C {\bf 73} (1997) 569.

\bibitem{Abreu:1995qx}
  P.~Abreu {\it et al.}  [DELPHI Collaboration],
  Z.\ Phys.\  C {\bf 67} (1995) 543.

\bibitem{Alexander:1996qi}
  G.~Alexander {\it et al.}  [OPAL Collaboration],
  Z.\ Phys.\  C {\bf 73} (1997) 587.

\bibitem{Acciarri:2000zf}
  M.~Acciarri {\it et al.}  [L3 Collaboration],
  Phys.\ Lett.\  B {\bf 479} (2000) 79.

\bibitem{Adam:1996hw}
  W.~Adam {\it et al.}  [DELPHI Collaboration],
  Z.\ Phys.\  C {\bf 70} (1996) 371.

\bibitem{Abreu:2000nu}
  P.~Abreu {\it et al.}  [DELPHI Collaboration],
  Phys.\ Lett.\  B {\bf 475} (2000) 429.

\bibitem{Abreu:1995we}
  P.~Abreu {\it et al.}  [DELPHI Collaboration],
  Phys.\ Lett.\  B {\bf 361} (1995) 207.

\bibitem{Alexander:1995gq}
  G.~Alexander {\it et al.}  [OPAL Collaboration],
  Phys.\ Lett.\  B {\bf 358} (1995) 162.

\bibitem{Abdallah:2003gu}
  J.~Abdallah {\it et al.}  [DELPHI Collaboration],
  Phys.\ Lett.\  B {\bf 569} (2003) 129.

\bibitem{Abreu:2000gw}
  P.~Abreu {\it et al.}  [DELPHI Collaboration],
  Eur.\ Phys.\ J.\  C {\bf 18} (2000) 203
  [Erratum-ibid.\  C {\bf 25} (2002) 493].



\end{thebibliography}
\end{document}